\documentclass[aps,pre,twocolumn,superscriptaddress,floatfix]{revtex4}
\usepackage{bm,graphicx,amsmath}

\newcommand{\E}[1]{Eq.~(\ref{#1})}
\newcommand{\F}[1]{FIG.~\ref{#1}}
\newcommand{\FS}[1]{FIGS.~\ref{#1}}
\newcommand{\T}[1]{TABLE~\ref{#1}}
\newcommand{\V}[1]{\vec{#1}}
\newcommand{\avg}[1]{\langle#1\rangle}
\begin{document}

\title{Intermittency in Two-Dimensional Turbulence with Drag}

\author{Yue-Kin~Tsang}

\email{yktsang@cims.nyu.edu}
\homepage{http://www.cims.nyu.edu/~yktsang}

\affiliation{Department of Physics, University of Maryland,
College Park, Maryland, 20742 USA}
\affiliation{Institute for Research in Electronics and Applied Physics,
University of Maryland, College Park, Maryland, 20742 USA}
\author{Edward~Ott}
\affiliation{Department of Physics, University of Maryland,
College Park, Maryland, 20742 USA}
\affiliation{Institute for Research in Electronics and Applied Physics,
University of Maryland, College Park, Maryland, 20742 USA}
\affiliation{Department of Electrical and Computer Engineering,
University of Maryland, College Park, Maryland, 20742 USA}
\author{Thomas~M.~Antonsen, Jr.}
\affiliation{Department of Physics, University of Maryland,
College Park, Maryland, 20742 USA}
\affiliation{Institute for Research in Electronics and Applied Physics,
University of Maryland, College Park, Maryland, 20742 USA}
\affiliation{Department of Electrical and Computer Engineering,
University of Maryland, College Park, Maryland, 20742 USA}
\author{Parvez~N.~Guzdar}
\affiliation{Institute for Research in Electronics and Applied Physics,
University of Maryland, College Park, Maryland, 20742 USA}

\date{\today}

\begin{abstract}
We consider the enstrophy cascade in forced two-dimensional turbulence with a
linear drag force. In the presence of linear drag, the energy wavenumber spectrum
drops with a power law faster than in the case without drag, and the vorticity
field becomes intermittent, as shown by the anomalous scaling of the vorticity
structure functions.  Using previous theory, we compare numerical simulation
results with predictions for the power law exponent of the energy wavenumber
spectrum and the scaling exponents of the vorticity structure functions
$\zeta_{2q}$. We also study, both by numerical experiment and theoretical
analysis, the multifractal structure of the viscous enstrophy dissipation in terms
of its R\'{e}nyi dimension spectrum $D_q$ and singularity spectrum $f(\alpha)$.
We derive a relation between $D_q$ and $\zeta_{2q}$, and discuss its relevance to
a version of the refined similarity hypothesis. In addition, we obtain and compare
theoretically and numerically derived results for the dependence on separation $r$
of the probability distribution of $\delta_{\V{r}}\omega$, the difference between
the vorticity at two points separated by a distance $r$. Our numerical simulations
are done on a $4096 \times 4096$ grid.
\end{abstract}



\maketitle

\section{Introduction}
Two-dimensional Navier-Stokes turbulence has attracted much interest because
of its relevance to a variety of natural flow phenomena. Examples are plasma
in the equatorial ionosphere \cite{kelley78} and the large-scale dynamics of
the Earth's atmosphere and oceans \cite{nastrom85}. In the laboratory, experiments
that are close to two-dimensional, such as soap film flow \cite{kellay98,rutgers98}
and magnetically forced stratified flow \cite{paret99}, have been conducted. In
addition, rotating fluid systems \cite{baroud02} are used to study quasi-two-dimensional
turbulence and its relevance to large-scale planetary flow. Ref.~\cite{kellay02} gives
a review of some recent experiments in two-dimensional turbulence.

In many of the situations involving two-dimensional turbulence, there are regimes where drag
is an important physical effect. In the ionospheric case, there is drag friction of the plasma
as it moves relative to the neutral gas background (due to ion-neutral collision). For geophysical
flows and rotating fluid experiments, viscosity and the no-slip boundary condition give rise to
Ekman friction. In this case the three-dimensional flow is often modeled as two-dimensional
outside the layer adjoining the no-slip boundary, and the effect of the boundary layer
manifests itself as drag in the two-dimensional description. For soap film and
magnetically forced flows, friction is exerted on the fluids by surrounding gas and
the bottom of the container, respectively. In all these cases, the drag force can be
modeled as proportional to the two-dimensional fluid velocity $\V{v}$, thus the
describing Navier-Stokes momentum equation becomes,
\begin{equation}
 \frac{\partial\V{v}}{\partial t}+(\V{v}\cdot\nabla)\V{v}
= -\frac{1}{\rho}\nabla p+\nu\nabla^2\V{v} -\mu\V{v}+\V{f}\ ,
\label{ns}
\end{equation}
where $\rho$ is the fluid density, $\mu$ is the drag coefficient, $\nu$ is the
kinematic viscosity and $\V{f}$ is an external forcing term. In two-dimensions, the
system can be described by the scalar vorticity field $\omega$ whose equation of motion
is obtained by taking the curl of \E{ns},
\begin{equation}
 \frac{\partial\omega}{\partial t}+\V{v}\cdot\nabla\omega
=\nu\nabla^2\omega-\mu\omega+f_\omega
\label{weq}
\end{equation}
with $\omega=\hat{z}\cdot\nabla\times\V{v}$ and $f_\omega=\hat{z}\cdot\nabla\times\V{f}$,
$\hat{z}$ being the unit vector perpendicular to the plane. In our studies, the forcing will
be taken to be localized at small wavenumbers ($k$) with a characteristic wavenumber $k_f$,
and incompressibility, $\nabla\cdot\V{v}=0$, will be assumed.

According to Kraichnan, for two-dimensional turbulence with no drag and very small viscosity,
for $k \gg k_f$, up to the viscous cutoff $k_d$, enstrophy cascades from small to large $k$
\cite{kraichnan67}. As a result, the energy wavenumber spectrum  $E(k)$ has a power law behavior
with logarithmic correction \cite{kraichnan71}, $E(k) \sim k^{-3}[\ln(k/k_f)]^{-1/3}$ for
$k_f \ll k \ll k_d$. In the presence of a linear drag, the energy spectrum drops with a power
law faster than the case with no drag \cite{nam00a,nam00b,bernard00},
\begin{equation}
E(k) \sim k^{-(3+\xi)} \qquad (\xi > 0)\ ,
\label{ek}
\end{equation}
and there is no logarithmic correction. A result similar to \E{ek} has been obtained for
the closely related problem of chaotically advected finite lifetime passive scalars
\cite{chertkov98,abraham98,nam99}. (See Section~\ref{sec:passive} for a discussion of
the relationship between vorticity in two-dimensional turbulence with drag and finite
lifetime passive scalars in Lagrangian chaotic flows.)

Furthermore, the vorticity field is intermittent as indicated by (i) anomalous
scaling of the {\it vorticity structure function}, (ii) scale dependence of the
{\it vorticity difference distribution} function, and (iii) multifractal
properties of the {\it enstrophy dissipation field}. Intermittency in the closely
related problem of chaotically advected finite lifetime passive scalars was
originally studied by Chertkov \cite{chertkov98} using a model flow in which
the velocity field was spatially linear and $\delta$-correlated in time (white
noise). This model successfully captures the generic intermittent aspect of the
problem \cite{note}. With respect to the situation of interest to us in
the present paper, it is {\it a priori} unclear to what extent the
$\delta$-correlated model yields results that approximate quantitative results for
intermittency measures obtained from experiments (numerical or real). This will be
discussed further in Secs.~\ref{sec:com} and \ref{sec:pdf}.

The structure function of order $2q$, $S_{2q}(\V{r})$ is defined as the
$(2q)^{\mathrm{th}}$ moment of the absolute value of the vorticity difference
$\delta_{\V{r}}\omega=\omega(\V{x}+\V{r})-\omega(\V{x})$. Assuming that the system
is homogeneous and isotropic, the structure functions depend on $r=|\V{r}|$ only.
For the case with drag ($\mu>0$), it is found that, in the enstrophy cascade range
$(k_d^{-1} \ll r \ll k_f^{-1})$, $S_{2q}(r)$ scales with $r$ as,
\begin{equation}
S_{2q}(r)=\avg{|\delta_{\V{r}}\omega|^{2q}} \sim r^{\zeta_{2q}}
\label{sqdef}
\end{equation}
with $\zeta_{2q} > 0$. Furthermore, the vorticity structure functions show anomalous scaling;
that is, $\zeta_{2q}$ is a nonlinear function of $q$. The nonlinear dependence of $\zeta_{2q}$
on $q$ indicates that the vorticity field is intermittent. In contrast, in the absence of
drag ($\mu=0$), it is predicted that $\omega(\V{x})$ wiggles
rapidly ({\it i.e.},
on the scale $k_d^{-1}$) and homogeneously in space. In terms of \E{sqdef}, this corresponds
to $\zeta_{2q}=0$ for $\mu=0$ \cite{falkovich94,eyink96,paret99}.

The intermittency of the vorticity field also manifests itself as a change in shape or form
of the probability distribution function of the vorticity difference $\delta_{\V{r}}\omega$
with the separating distance $r$. It can be shown that if the probability distribution function
$\bar{P}_r(X_{\V{r}})$ of the standardized vorticity difference,
\begin{equation}
X_{\V{r}}
=\frac{\delta_{\V{r}}\omega}{\sqrt{\avg{(\delta_{\V{r}}\omega)^2}}}\ ,
\label{xr}
\end{equation}
is independent of $r$, then $\zeta_{2q}$ increases linearly with $q$: From \E{sqdef} and \E{xr},
\begin{equation}
S_{2q}(r)=\avg{|\delta_{\V{r}}\omega|^2}^q\int|X_{\V{r}}|^{2q}\bar{P}_r(X_{\V{r}})\,dX_{\V{r}}\ ,
\label{s2qpx}
\end{equation}
and, if $\bar{P}_r(X_{\V{r}})$ is independent of $r$, for $r$ in the cascade range, then the
only $r$-dependence of $S_{2q}(r)$ comes through the term $\avg{|\delta_{\V{r}}\omega|^2}^q$,
implying $\zeta_{2q}=q\zeta_2$; that is, $\zeta_{2q}$ is linearly proportional to $q$. Such
collapse of $\bar{P}_r(X_{\V{r}})$ for different values of $r$ has been observed in an experiment
\cite{paret99} where drag is believed to be unimportant. When the effect
of drag is not negligible, $\bar{P}_r(X_{\V{r}})$ changes shape and develops exponential
or stretched-exponential tails as $r$ decreases, so that \E{s2qpx} admits nonlinear dependence
of $\zeta_{2q}$ on $q$.

The intermittency of the vorticity field is closely related to the multifractal
structure of the viscous enstrophy dissipation field. We define the enstrophy as $\omega^2/2$.
From \E{weq}, the time evolution of the enstrophy is then given by
\begin{eqnarray}
\frac{\partial}{\partial t}\left(\frac{\omega^2}{2}\right)
&=& -\nabla\cdot\left[\V{v}\left(\frac{\omega^2}{2}\right)
    -\nu\nabla\!\!\left(\frac{\omega^2}{2}\right)\right]-\nu|\nabla\omega|^2 \nonumber \\*
&&-2\mu\left(\frac{\omega^2}{2}\right)+\omega f_\omega\ .
\label{densdt}
\end{eqnarray}
We identify the second term on the right hand side of \E{densdt} as the local rate of
viscous enstrophy dissipation $\eta$,
\begin{equation}
\eta=\nu|\nabla\omega|^2\ .
\label{eta}
\end{equation}
The multifractality of the viscous enstrophy dissipation can be quantified by the R\'{e}nyi
dimension spectrum of a measure based on the vorticity gradient. Imagine we divide
the region $\mathcal{R}$ occupied by the fluid into a grid of square boxes of size $\epsilon$,
we define the measure $p_i$ of the $i^\mathrm{th}$ box $\mathcal{R}_i(\epsilon)$ as
\begin{equation}
p_i(\epsilon)
=\frac{\int_{\mathcal{R}_i(\epsilon)}|\nabla\omega|^2\,d\V{x}}
{\int_\mathcal{R}|\nabla\omega|^2\,d\V{x}}\ .
\label{pidef}
\end{equation}
The R\'{e}nyi dimension spectrum \cite{renyi70} based on this measure is then given by,
\begin{equation}
D_q=\frac{1}{q-1}\lim_{\epsilon\rightarrow 0}\lim_{\nu\rightarrow 0}
    \frac{\log\sum_i p_i^q}{\log\epsilon}\ .
\label{renyidq}
\end{equation}
The definition \E{renyidq} was introduced in the context of natural measures occurring
in dynamical systems by Grassberger \cite{grassberger83}, and Hentschel and Procaccia
\cite{hentschel83}. In the case with drag, we find that the dimension spectrum for $p_i$
is multifractal; that is, $D_q$ varies with $q$, in contrast to the case with no drag in
which $D_q=2$, indicating that the measure is uniformly rough.

The measure $p_i$ can also be described in terms of the singularity spectrum $f(\alpha)$
\cite{halsey86}. In particular, to each box $\mathcal{R}_i$, we associate a singularity index
$\alpha_i$ via
\begin{equation}
\alpha_i=\frac{\log p_i}{\log\epsilon}\ ,
\label{pi}
\end{equation}
and the number of boxes $N(\alpha)d\alpha$ with singularity index between $\alpha$ and
$\alpha+d\alpha$ is then assumed to scale as
\begin{equation}
N(\alpha)\sim\epsilon^{-f(\alpha)}\ .
\label{na}
\end{equation}
$f(\alpha)$ can loosely be interpreted as the dimension of the set of boxes with
singularity index $\alpha$ \cite{ott02}. When $f(\alpha)$ and $D_q$ are smooth functions,
$f(\alpha)$ is related to $D_q$ by a Legendre transformation \cite{halsey86}. The multifractal
nature of the viscous enstrophy dissipation in the presence of drag implies that the $f(\alpha)$
spectrum, defined by \E{pi} and \E{na}, is a nontrivial function of $\alpha$.

The subject of this paper is the relation of $\bar{P}_r(X_{\V{r}})$, the exponents, $\xi$ and
$\zeta_{2q}$, and the fractal dimension $D_q$ to the chaotic properties of the turbulent flow.
In chaotic flows, the infinitesimal separation between two fluid particle trajectories,
$\delta\V{x}(t)$ typically diverges exponentially. The net rate of exponentiation
over a time interval from 0 to $t$ for a trajectory starting at $\V{x}_0$ is
given by the finite-time Lyapunov exponent, $h$ defined as,
\begin{equation}
h(t;\V{x}_0)=\frac{1}{t}\log\frac{|\delta\V{x}(t)|}{|\delta\V{x}(0)|}\,.
\label{h}
\end{equation}
At a particular time $t$, $h(t;\V{x}_0)$ in general depends on the initial
positions $\V{x}_0$ and the initial orientation of the perturbation $\delta\V{x}(0)$. 
However, for large $t$, the results for $h(t;\V{x}_0)$ is insensitive to the orientation
of $\delta\V{x}(0)$ for typical choices of $\delta\V{x}(0)$, and we neglect the dependence
on $\delta\V{x}(0)$ in what follows. The distribution in the values of $h$ for randomly
chosen $\V{x}_0$ can be characterized by the conditional probability density function $P(h\,|\,t)$.
In subsequent sections, we shall briefly review the theories that relate $\xi$ \cite{nam00a} and
$\zeta_{2q}$ \cite{neufeld00} to the distribution $P(h\,|\,t)$ and the drag coefficient $\mu$. We
then derive an expression for $\bar{P}_r(X_{\V{r}})$ and a relation between $D_q$ and $\zeta_{2q}$.
We apply these theories to compute $\xi$, $\zeta_{2q}$, $\bar{P}_r(X_{\V{r}})$ and $D_q$ for turbulent
flows governed by \E{weq} and the theoretical results are compared to those obtained from direct
numerical simulations.

We perform our simulations on a square domain of size
$[-\pi,\pi] \times [-\pi,\pi]$ with periodic boundary conditions in both
directions. The viscous term in \E{weq} is replaced by a hyperviscous damping
$-\nu\nabla^8\omega$ with $\nu=7.5\times10^{-25}$ and $f_\omega(x,y)=\cos 2x$ is
used for the source function of the vorticity. Our use of a hyperviscosity is similar
to what is often done in numerical studies of three-dimensional turbulence and, for a
given numerical resolution, has the desirable effect of increasing the scaling range
where dissipation can be neglected, while, at the same time it is hoped that this
change in the dissipation does not influence the scaling range physics. For wavenumber
$k \leq 6$, we have $\mu=0.1$ and this provides an energy sink at the large scales.
For $k > 6$, we will consider the cases of $\mu=0.1$ and $\mu=0.2$. As we shall see in
Section~\ref{sec:passive}, when drag is present, the large $k$ vorticity components can be
considered as being passively advected by the small-$k$ flow components. Applying the same
small-$k$ drag ({\it i.e.}, $\mu$=0.1 for $k\leq 6$) allows us to compare the effects of drag
at small scales while keeping similar large scale dynamics of the flows. For all the numerical
results presented here, we use a spatial grid of $4096\times 4096$ and a time step of
0.00025. Starting from random initial conditions for the vorticity field, \E{weq} is integrated
using a time split-step method described in detail in Ref.~\cite{yuan00}. The system appears to
reach a statistical steady-state after about 40 time units. \F{snap} shows snapshots of the vorticity
field for the cases of $\mu(k>6)=0.1$ and $\mu(k>6)=0.2$.%
\begin{figure}
\includegraphics[scale=0.415]{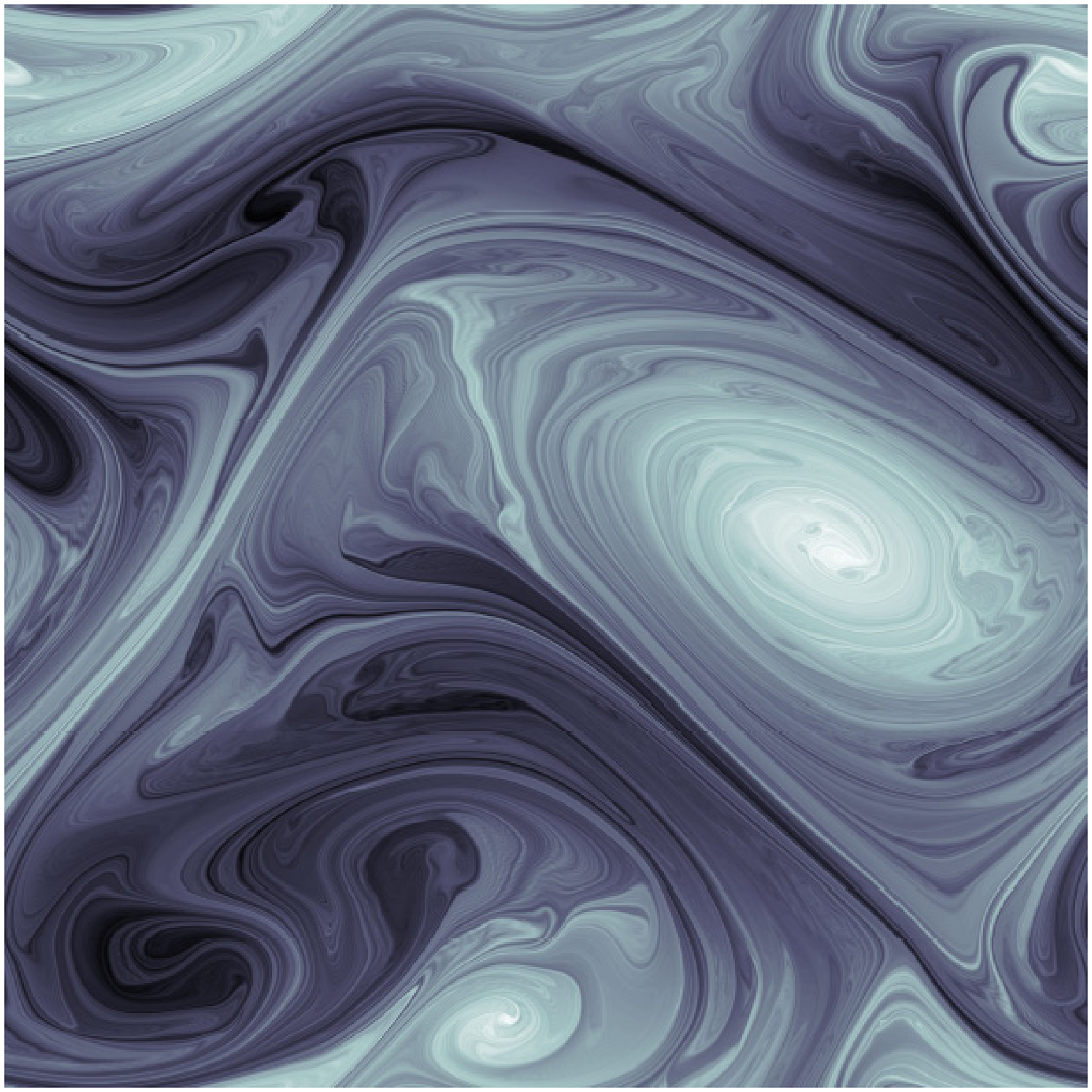}\\%
\vspace{0.3cm}
\includegraphics[scale=0.415]{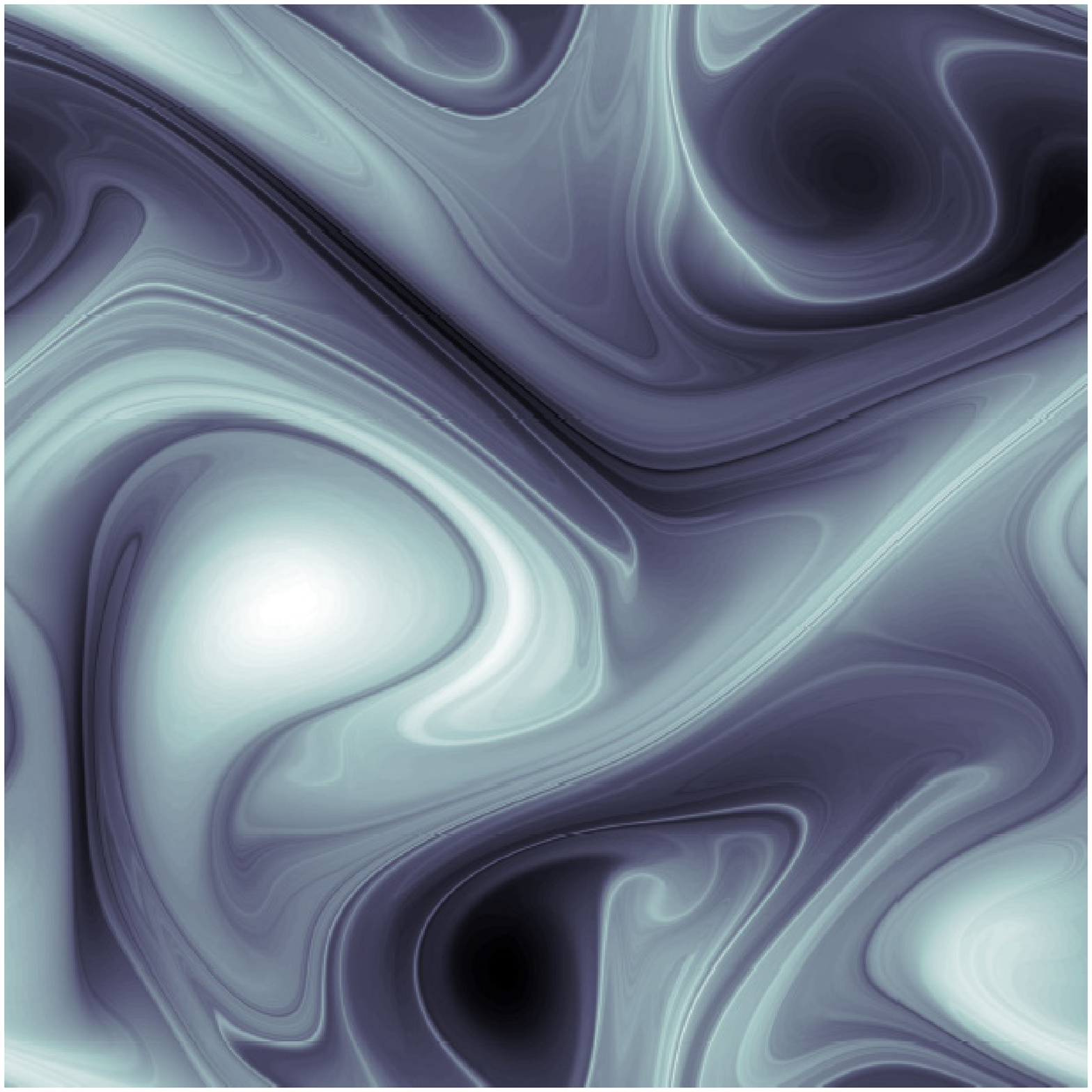}%
\caption{Snapshots of the vorticity field $\omega(x,y)$ at $t=61$ for the case
$\mu(k>6)=0.1$ (upper) and at $t=65$ for the case $\mu(k>6)=0.2$ (lower) ($\mu=0.1$
for $k\leq 6$ for both cases). Light areas are regions of large positive values of the
vorticity, and dark areas are regions of negative vorticity of large magnitude. The scales
used in the two plots are not the same.}
\label{snap}
\end{figure}
We note that the vorticity field for the case with a larger drag shows less fine structure
and the large scale vortices tend to have shorter lifetime. At any given moment, there are
typically 3 -- 5 large vortex structures visible in the system. In the steady-state, vortices
are continuously created and destroyed.

\section{Role of the finite-time Lyapunov exponent}
\label{sec:wpmpht}

\subsection{Passive Nature of the Small-Scale Vorticity}
\label{sec:passive}

The theory that we will use is based on the approximation that the high $k$ components of the
vorticity field are passively advected by the large scale structures of the flow. This can be
justified by the following argument given in Ref.~\cite{nam00a}. The Lyapunov exponent $h$ of the
flow is the mean rate of exponentiation of differential displacement $\delta\V{x}$ following the
flow, where $d(\delta\V{x})/dt=\delta\V{x}\cdot\nabla\V{v}$. Thus one can crudely estimate the
Lyapunov exponent as
\begin{equation}
h \sim \avg{\|\nabla\V{v}\|^2}^{\frac{1}{2}} \sim \sqrt{\int_{k_f}^\infty k^2E(k)dk}\ ,
\label{happrox}
\end{equation}
where $\|\nabla\V{v}\|^2=(\partial v_x/\partial x)^2+(\partial v_y/\partial
y)^2+(\partial v_x/\partial y)^2+(\partial v_y/\partial x)^2$.  Assuming the limit
of infinite Reynolds number and power law behavior of $E(k)$ valid as
$k\rightarrow\infty$, the integral in \E{happrox} diverges at the upper limit
unless $\xi$ in \E{ek} is positive. That is, the velocity field $\V{v}$ is not
differentiable ($h$ and $\nabla\V{v}$ are undefined) unless $\xi>0$
(alternatively, if $\xi<0$ and viscosity imposes a cutoff to power law behavior of
$E(k)$ at $k\sim k_d$, then, although $\|\nabla\V{v}\|^2$ is now finite, the
integral in \E{happrox} is dominated by velocity components at the shortest
wavelength).  From \E{happrox}, for $\xi>0$, we have $h\sim k_f^{-\xi/2}$. This
means that $h$, which characterizes small scale stretching, is determined by the
largest scale flow components. Since $\xi>0$ in the case where drag is present,
$\nabla\V{v}$ is predominantly determined by the largest spatial scales. Thus the
Lyapunov exponents provide information on the evolution of the distance between
fluid elements whose separation is {\it finite} but small compared to
$k_f^{-1}$. This will allow us to approximate the evolution of vorticity field
components with wavenumbers in the range $k_f \ll k < k_d$ using Lyapunov
exponents that result primarily from the large spatial scales $k\sim k_f$. That
is, the vorticity field at wavenumber $k_f \ll k < k_d$ evolves via approximately
passive advection by the large scale flow.  (Note that for $\xi<0$ such an
approach would not be applicable since the Lyapunov numbers would provide an
estimate of separation evolution only for distances less that $k_d^{-1}$ which is
outside the dissipationless power law range.)

Note that the case without drag corresponds to $\xi=0$, which is marginal in the
sense that it is on the borderline of the condition for differentiability of the
velocity field. In other situations of marginality (e.g., in critical phenomena),
it is often found that there are logarithmic corrections to power-law scaling, and
this may be thought of as the origin of Kraichnan's logarithmic correction to the
$k^{-3}$ enstrophy cascade scaling of $E(k)$.

\subsection{Distribution of Finite-time Lyapunov Exponent}
\label{sec:pht}

As mentioned earlier, the exponential divergence of nearby trajectories in a
chaotic flow over a time interval $0$ to $t$ can be quantified by a finite-time
Lyapunov exponent $h(t;\V{x}_0)$ defined in \E{h}. In the limit
$t\rightarrow\infty$, $h(t;\V{x}_0)$ will approach the usual infinite-time
Lyapunov exponent $\bar{h}$ for almost all initial conditions $\V{x}_0$ and almost
all initial orientations of $\delta\V{x}$.  At finite time, the dependence of $h$
on $\V{x}_0$ results in a distribution in the values of $h$ which can be
characterized by the probability density function $P(h\,|\,t)$. That is, if
$\V{x}_0$ is chosen randomly with uniform density in the region of the fluid flow,
and if the initial orientation of $\delta\V{x}$ is arbitrarily chosen, then we can
define a probability distribution $P(h\,|\,t)$ such that $P(h\,|\,t)dh$ is the
probability that $h(t;\V{x}_0)$ lies between $h$ and $h+dh$. As $t$ increases,
$P(h\,|\,t)$ will become more and more sharply peaked at $\bar{h}$ and approach a
delta-function at $\bar{h}$ as $t\rightarrow\infty$.

The passive nature of the small-scale flow allows us to use the procedures described in
Ref.~\cite{yuan00} to obtain histogram approximations to $P(h\,|\,t)$. \F{pht} shows $P(h\,|\,t)$
at different $t$ for the case of $\mu=0.2$.%
\begin{figure}
\includegraphics[scale=0.4]{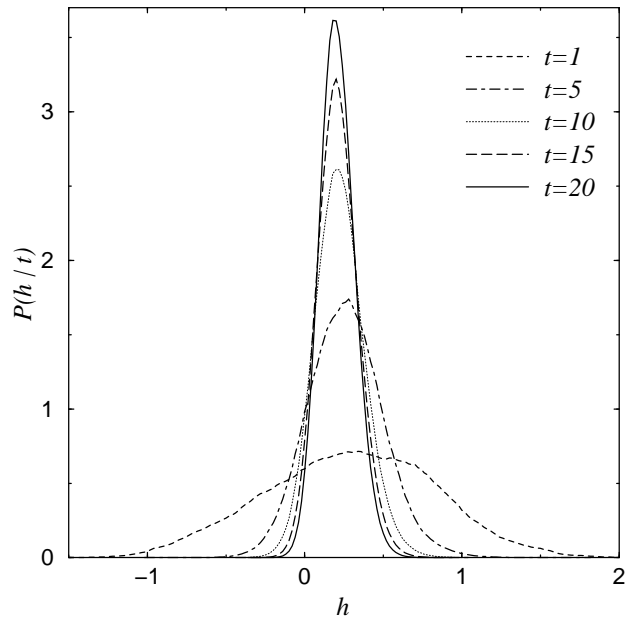}%
\caption{Probability distribution function of the finite-time Lyapunov exponent
$P(h\,|\,t)$ for different time $t$ for the case $\mu=0.2$.}
\label{pht}
\end{figure}
As time increases, $P(h\,|\,t)$ becomes sharply peak around a positive value of $h$
showing that the flow is chaotic. In particular, for $t=20$, $P(h\,|\,t)$ has its peak
at $\bar{h}\approx 0.20$. The function $P(h\,|\,t)$ shows similar behavior for the case $\mu=0.1$
with a peak occurring at $\bar{h}\approx 0.24$ for $t=20$. The smaller damping in the latter
case apparently allows the fluid to undergo bigger stretching in a given amount of time.

Based on the argument that $h(t;\V{x}_0)$ can be considered as an average over
many independent random realizations, $P(h\,|\,t)$ is approximated by the following
asymptotic form (\cite{ott02} and references therein),
\begin{equation}
P(h\,|\,t)\sim\sqrt{\frac{tG''(h)}{2\pi}}\,e^{-tG(h)}
\label{gh}
\end{equation}
for large $t$. \E{gh} has been shown to be true for the generalized baker's map \cite{ott89}
and numerically verified for cases where there are no KAM surfaces, for example, see
Ref.~\cite{varosi91}. The function $G(h)$ is concave upward, $G''(h)\geq0$. It has the
minimum value zero, occurring at $h=\bar{h}$, $G'(\bar{h})=G(\bar{h})=0$. The significance
of \E{gh} is that it expresses a function of two variables $P(h\,|\,t)$ in terms of a function
of a single variable $G(h)$. \E{gh} will be used in the development of the theory. We note that
$G(h)$ is completely specified by the flow $\V{v}(\V{x},t)$ and hence, is dependent on the
value of $\mu$.

The function $G(h)$ can be approximated at large $t$ from $P(h\,|\,t)$ using the following
relation,
\begin{equation}
G(h)\approx K-\frac{1}{t}\ln P(h\,|\,t)\ ,
\label{ghapp}
\end{equation}
where $K$ is determined by the condition that the minimum of $G(h)$ equals zero. \F{gh1}
shows the $G(h)$ obtained from the corresponding $P(h\,|\,t)$ shown in \F{pht}.%
\begin{figure}
\includegraphics[scale=0.4]{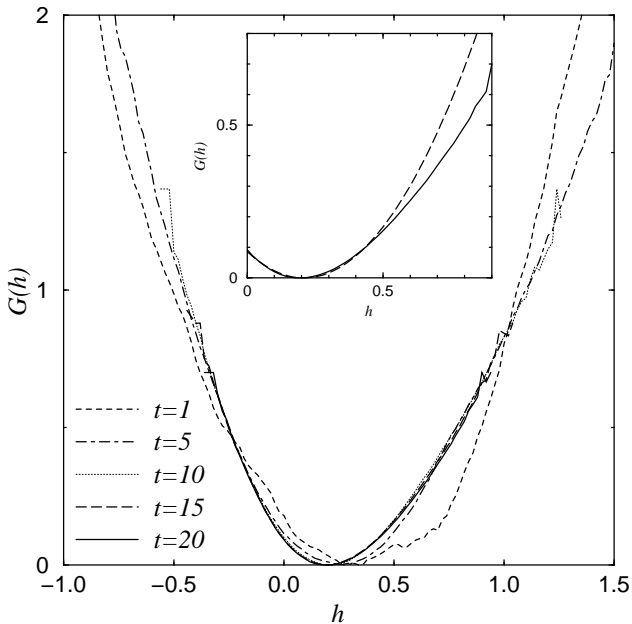}%
\caption{$G(h)$ (see \E{gh}) for different time $t$ for the case $\mu(k>6)=0.2$. The
inset shows $G(h)$ obtained at $t=20$ (solid line) and the quadratic
approximation to $G(h)$ obtained by the method described at the end of
Sec.~\ref{sec:com} (dashed line).}
\label{gh1}
\end{figure}
The large number $(4096^2)$ of fluid trajectories used in the generation of
each $P(h\,|\,t)$ allows us to obtain $G(h)$ for a large range of $h$. As can be seen in
\F{gh1}, for large enough $t$, the graphs of $G(h)$ for different $t$ more or less
collapse onto each other, showing that \E{gh} is a good approximation to $P(h\,|\,t)$
for the flows we considered. $G(h)$ shows similar behavior in the case $\mu=0.1$.

\section{The Exponents $\xi$ and $\zeta_{2q}$}
\label{sec:xi_zeta}

\subsection{Review of Theory}
\label{sec:ek_sq_th}

We consider the scaling of the vorticity structure functions in the limit $\nu\rightarrow 0^+$.
The results have previously been given in Ref.~\cite{neufeld00,chertkov98}, which treats
finite-lifetime passive scalars \cite{abraham98,chertkov98,nam99} rather than vorticity in
two-dimensional turbulence with drag, and in Ref.~\cite{boffetta02} for two-dimensional turbulence
with drag. Here, we re-derive the previous results (our derivation is different from that in
Ref.~\cite{neufeld00,chertkov98} and more complete than that in Ref.~\cite{boffetta02}).

From \E{weq} with initial condition $\omega(\V{x},0)=0$, we have \cite{neufeld00},
\begin{eqnarray}
\label{deltaw}
\delta_{\V{r}}\omega(\V{x},t)
&=& \int_0^t\delta f_\omega[\V{x}(t')]e^{\mu(t-t')}dt' \nonumber \\*
&\sim&\int_{t-\tau(r)}^t\V{r}\cdot\nabla f_\omega[\V{x}(t')]e^{(\mu-h)(t'-t)}dt' \nonumber \\*
&&+\ \int_0^{t-\tau(r)}\delta f_\omega[\V{x}(t')]e^{\mu(t'-t)}dt'\ .
\end{eqnarray}
where $\delta f_\omega[\V{x}(t')]=f_\omega[\V{x}(t')+\delta\V{x}(t')]-f_\omega[\V{x}(t')]$,
$\V{x}(t')$ and $\V{x}(t')+\delta\V{x}(t')$ are the two trajectories that pass through
$\V{x}$ and $\V{x}+\V{r}$ at $t'=t$; thus $\delta\V{x}(t)\equiv\V{r}$. Since the points
$\V{x}$ and $\V{x}+\V{r}$ are specified at
time $t$, chaos in the flow implies that the backward evolved trajectories to times $t'<t$ diverge
exponentially from each other, $\delta\V{x}(t')\sim\V{r}e^{h(t-t')}$, until $|\delta\V{x}(t')|$
reaches the system size $L$, past which $|\delta\V{x}(t')|$ remains of order $L$. We define $\tau(r)$
such that $|\delta\V{x}(t-\tau)|\sim L \sim re^{h\tau}$. We have linearized $\delta f_\omega[\V{x}(t')]$
for $t-\tau(r)<t'<t$ when the two trajectories are close together. On the other hand, for
$0<t'<t-\tau(r)$, $|\delta\V{x}(t')|\sim L$ and $\delta f_\omega[\V{x}(t')]$ thus fluctuates with
roughly constant amplitude, $\delta f_\omega \sim f_\omega$. Thus the integral from 0 to $t-\tau(r)$
in \E{deltaw} is of the order of $e^{-\mu\tau}$ \cite{boffetta02}. The integral from $t-\tau(r)$
to $t$ in \E{deltaw} shows the same $\sim e^{-\mu\tau}$ behavior if $h>\mu$ but is 
of the order of $e^{-h\tau}$ if $h<\mu$. Hence, using the definition of $S_{2q}(r)$ and replacing
the average over $\V{x}$ by an average over $\tau$ at fixed exponentiation $\lambda=\ln(L/r)=h\tau$,
we find
\begin{equation}
\label{avgt}
S_{2q}(r) \sim
\int_0^{\frac{\lambda}{\mu}}\!d\tau\,R(\tau\,|\,\lambda)e^{-2q\mu\tau}+
\int_{\frac{\lambda}{\mu}}^\infty\!d\tau\,R(\tau\,|\,\lambda)e^{-2q\lambda}\ .
\end{equation}
The conditional probability density $R(\tau\,|\,\lambda)$ of $\tau$ for a fixed $\lambda$ is
related to $P(h\,|\,\tau)$ by \cite{reyl98}
\begin{equation}
R(\tau\,|\,\lambda) \approx \frac{d}{d\tau}\int_{\frac{\lambda}{\tau}}^\infty\!dh\,P(h\,|\,\tau)\ .
\label{rtlapp}
\end{equation}
Using the asymptotic form \E{gh} for $P(h\,|\,\tau)$, the integral \E{avgt} is performed using the
steepest descent method with $\lambda$ as the large parameter.
Thus we obtain that the structure function scaling exponents, $\zeta_{2q}$ defined in \E{sqdef},
is given by
\begin{equation}
\zeta_{2q}=
\displaystyle \min_h\{2q,H_q(h)\}\ ,
\label{zeta2q}
\end{equation}
where
\begin{equation}
H_q(h)=\frac{G(h)+2q\mu}{h}\ .
\label{hq}
\end{equation}
Equations (\ref{zeta2q}) and (\ref{hq}) yield the previous passive scalar result
of Ref.~\cite{chertkov98} if $G(h)$ is assumed parabolic,
$G(h)=(constant)(h-\bar{h})^2$, which is a consequence of the temporally white
noise velocity field model of Ref.~\cite{chertkov98}.

We now consider the energy density $e(\V{k})$ which is given by
\begin{equation}
e(\V{k})=\frac{1}{(2\pi)^2}\frac{|\tilde{v}_x(\V{k})|^2+|\tilde{v}_y(\V{k})|^2}{2L^2}\,,
\end{equation}
where $\tilde{v}_x(\V{k})$ and $\tilde{v}_y(\V{k})$ are Fourier transforms of the $x$
and $y$ components of the velocity field $\V{v}(x,y)$. The wavenumber energy spectrum
$E(k)$ is then defined as
\begin{equation}
E(k)=\int d\V{k}'\delta(|\V{k}'|-k)e(\V{k}')\,.
\label{ekdef}
\end{equation}
A previous theory \cite{nam00a,nam00b} relates the energy wavenumber spectrum exponent,
$\xi$ defined in \E{ek} to $G(h)$. The result is
\begin{equation}
\xi=\min_h\{H_1(h)\}\ .
\label{xi}
\end{equation}
Thus, $\xi$ and $\zeta_{2q}$ are related to the properties of the flow, namely the drag coefficient
$\mu$ and the distribution of the finite-time Lyapunov exponent $h$.

Let $h=h_q^*$ be the value of $h$ at which $H_q(h)$ is minimum. We now show that
$h_q^*$ increases with $q$. Letting $\beta > \alpha$, by the definition of $h_q^*$,
$H'_\alpha(h_\alpha^*)=H'_\beta(h_\beta^*)=0$, it follows that $G'(h_\beta^*)-H_\beta(h_\beta^*)
=G'(h_\alpha^*)-H_\alpha(h_\alpha^*)$. Since by \E{hq}, $H_\beta(h)>H_\alpha(h)$ for all $h$, we
have $G'(h_\beta^*)>G'(h_\alpha^*)$ which implies $h_\beta^*>h_\alpha^*$ due to the fact that 
$G''(h)>0$ for all $h$. Moreover, putting $\alpha=0$ gives $h_q^*>\bar{h}$ for all $q$.

\subsection{Comparison of Theory and Numerical Results}
\label{sec:com}

To apply the theory \E{zeta2q} and \E{xi} to numerically determine the exponents $\zeta_{2q}$ and
$\xi$ of our turbulent flow governed by \E{weq}, we first let
\begin{equation}
\bar\zeta_{2q}=\min_h\{H_q(h)\}\ .
\label{tzeta}
\end{equation}
Using \E{hq}, we re-write \E{tzeta} as
\begin{equation}
\min_h\left\{\frac{G(h)+2q\mu}{h}-\bar\zeta_{2q}\right\}=0\ .
\label{minh}
\end{equation}
Since the function minimized in \E{minh} has minimum value zero, we can multiply it by any positive
function of $h$, and the minimum will still be zero. Thus for $h>0$ we can multiply the minimized
function by $h$ to obtain
\begin{equation}
\min_h\{G(h)-h\bar\zeta_{2q}\}=-2q\mu\ .
\label{minhqm}
\end{equation}
Using \E{gh} and \E{minhqm}, we see that steepest descent evaluation of the following integral
for large $t$ yields
\begin{equation}
\int P(h\,|\,t) e^{h\bar\zeta_{2q}t}\,dh \sim e^{2q\mu t}\ .
\label{steep}
\end{equation}
Thus we define the partition function \cite{reyl98}
\begin{equation}
\Gamma(z,t)=\int e^{zh(t;\V{x}_0)t}\,d\V{x}_0\ ,
\label{gzt}
\end{equation}
in terms of which \E{steep} becomes
\begin{equation}
\Gamma(\bar\zeta_{2q},t)\sim e^{2q\mu t}\ .
\end{equation}
We numerically compute the partition function \E{gzt} using the approximation,
\begin{equation}
\Gamma(z,t)=\frac{1}{M}\sum_{i=1}^M\,e^{zh(t;\V{x}_{0i})t}
\label{gammazt}
\end{equation}
employing many ($M$=$4096^2$) spatially uniformly distributed initial conditions $\V{x}_{0i}$
$(i=1,2,\ldots, M)$.

For different values of $\bar\zeta_{2q}$, we then plot $\ln\Gamma(\bar\zeta_{2q},t)$ versus
$t$. \F{ehzt} shows samples of $\ln\Gamma(\bar\zeta_{2q},t)$ for the case $\mu=0.1$.%
\begin{figure}
\includegraphics[scale=0.4]{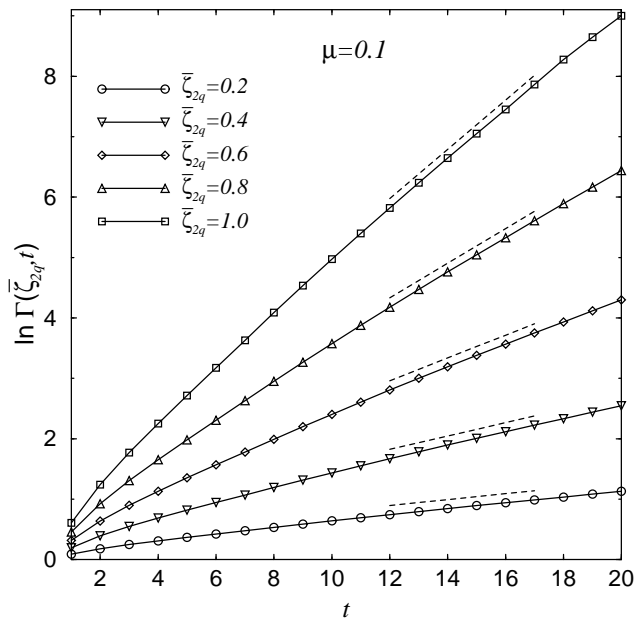}%
\caption{Log-linear plot of the partition function $\Gamma(\bar\zeta_{2q},t)$ for $\mu=0.1$.
The dotted lines are linear fits.}
\label{ehzt}
\end{figure}
As expected, for large $t$, $\ln\Gamma(\bar\zeta_{2q},t)$ is linear in $t$ and the
slope, which can be estimated using a linear fit, gives the corresponding value of
$q$ for each $\bar\zeta_{2q}$.  \F{z2q_ehzt} plots $\bar\zeta_{2q}$ versus $q$
obtained in this way for $\mu=0.1$ and $\mu(k>6)=0.2$ (open circles and squares)
together with the corresponding sixth degree polynomial fits (solid lines).%
\begin{figure}
\includegraphics[scale=0.4]{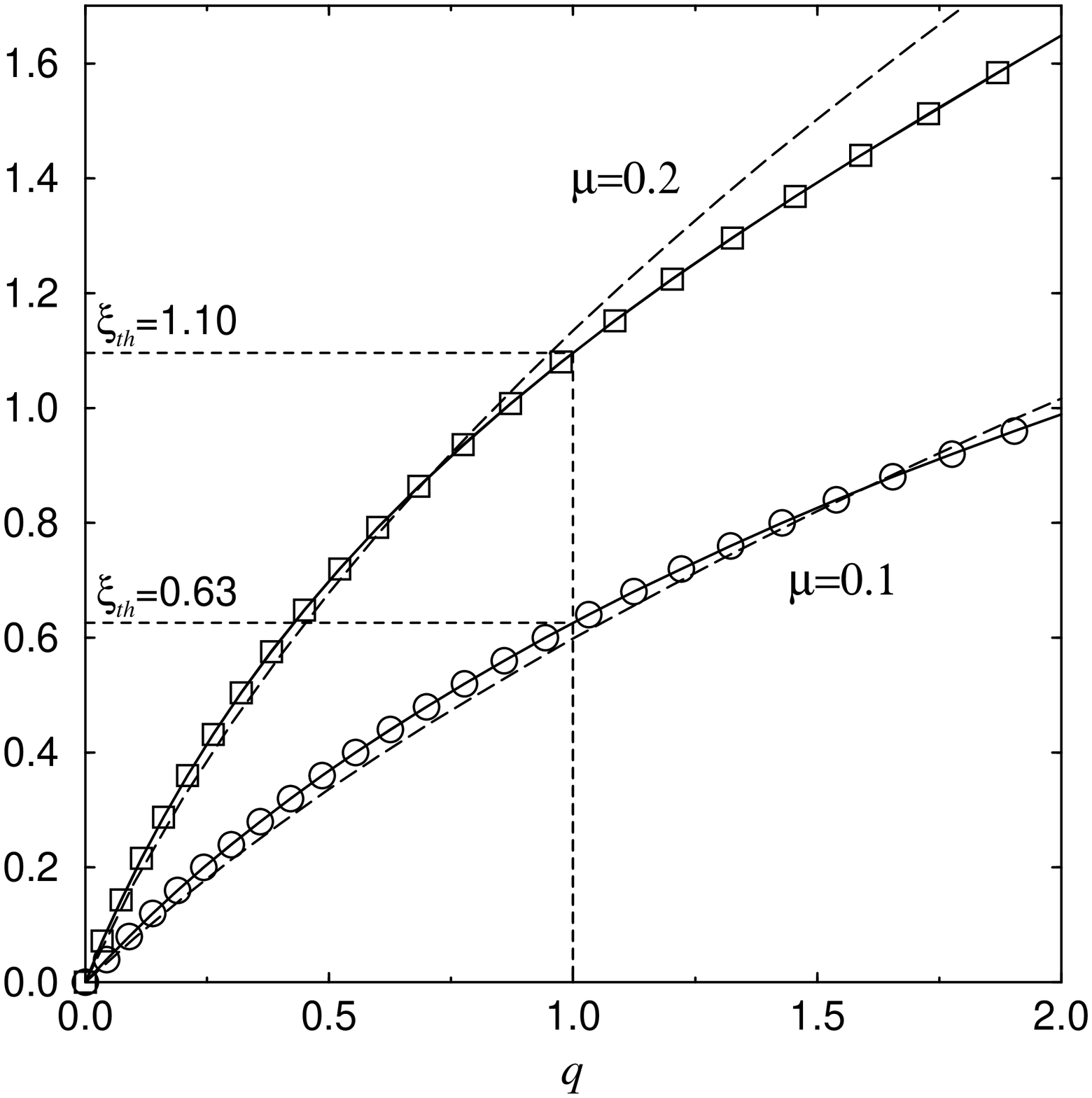}%
\caption{$\bar\zeta_{2q}$ as a function of $q$ from the method of \F{ehzt} for
$\mu=0.1$ (circle) and $\mu(k>6)=0.2$ (square). The solid lines are sixth degree
polynomial fits to circles and squares. Results using Eq.~(5) of
Ref.~\cite{chertkov98} are shown as dashed lines.}
\label{z2q_ehzt}
\end{figure}
By \E{xi} and \E{zeta2q}, the exponents $\xi$ and $\zeta_{2q}$ can then be determined from
\F{z2q_ehzt} by $\xi=\bar\zeta_{2}$ and $\zeta_{2q}=\min\{2q,\bar\zeta_{2q}\}$.

It is also possible to compute $\xi$ and $\zeta_{2q}$ directly from \E{xi} and
\E{zeta2q} using a fourth degree polynomial fit of the numerically obtained
$G(h)$. We find that the two methods give similar results, but that the method of
\E{gammazt} is easier to compute.

We also test the applicability of the result of Ref.~\cite{chertkov98}, which
employs a model in which the velocity field is $\delta$-correlated in time. This
model may be shown to correspond to \E{zeta2q} and \E{hq} with $G(h)$ parabolic,
$G(h)=a(h-\bar{h})^2$ where $\bar{h}$ is the infinite time Lyapunov exponent. With
this form of $G(h)$, an explicit analytic expression for the structure function
exponents can be obtained in terms of the constant $a$ and $\bar{h}$; see Eq.~(5)
of Ref.~\cite{chertkov98}. In order to apply this result to a specific flow, we
need to formulate a procedure for obtaining a reasonable value of $a$ from the flow
($\bar{h}$ is well-defined and numerically accessible by standard technique). For
this purpose, we note that, within the $\delta$-correlated approximation, the
total exponentiation $h(t,\V{x}_0)t$ experienced by an infinitesimal vector
originating at position $\V{x}_0$ undergoes a random walk with diffusion constant
$D=a/4$. Thus we obtain $a$ as one half of the long time slope of a plot of
$\avg{(h-\bar{h})^2}t^2$ versus $t$. The inset to \F{gh1} shows as a solid curve
$G(h)$ obtained using \E{ghapp} (as previously described) along with
the parabolic model (dashed curve) with $a$ determined by the above
procedure. There appears to be a substantial difference in the important range
$h>\bar{h}$ where the saddle points are located. Somewhat surprisingly, however,
this does not lead to much difference in the numerical values of the structure
function exponents. This is shown in \F{z2q_ehzt} where the results using Eq.~(5)
of Ref.~\cite{chertkov98} are plotted as dashed curves. In fact, the difference is
within the error of our numerical experiments that directly determine the values
of $\zeta_{2q}$. Thus for this case the parabolic approximation provides an
adequate fit to the data, although this could not be definitely predicted {\it a priori}.

\subsubsection{Energy Wavenumber Spectrum}
\label{sec:ek_com}

As illustrated in \F{z2q_ehzt}, theoretical predictions for $\xi$, denoted as $\xi_{th}$, are
obtained using the method described above. The results are given in \T{tbl:xi}.%
\begin{table}[b]
\caption{Comparison of the values of $\xi$ obtained from numerical simulations to
the theoretical results.}
\label{tbl:xi}
\begin{ruledtabular}
\begin{tabular}{ccccc}
$\mu(k>6)$ & $\xi_{th}(=\zeta_{2,th})$ & $\xi_{DNS}$ & $\zeta_{2,DNS}$ & $\zeta_{2,E(k)}$\\ \hline
0.1        & 0.63                        & 0.61        & 0.66          & 0.68\\
0.2        & 1.10                        & 1.12        & 1.16          & 1.14
\end{tabular}
\end{ruledtabular}
\end{table}
To verify the theoretical results, we compute the energy spectrum directly from the
numerical solution of \E{weq} on a $4096\times 4096$ grid using \E{ekdef}, which can be 
written in terms of the vorticity as
\begin{equation}
E(k,t)=\int\frac{d\V{k}'}{(2\pi L)^2}\,
\delta(|\V{k}'|-k)\frac{|\tilde{\omega}(\V{k}',t)|^2}{2|\V{k}'|^2}
\end{equation}
where $\tilde{\omega}(\V{k}',t)$ is the Fourier transform of $\omega(\V{x},t)$. The time
averaged energy spectrum $E(k)$ is obtained by averaging $E(k,t)$ at every 0.1 time unit
from $t=41$ to $t=75$. \F{fig:ek} shows a log-log plot of $E(k)$ versus $k$ for the two%
\begin{figure}
\includegraphics[scale=0.4]{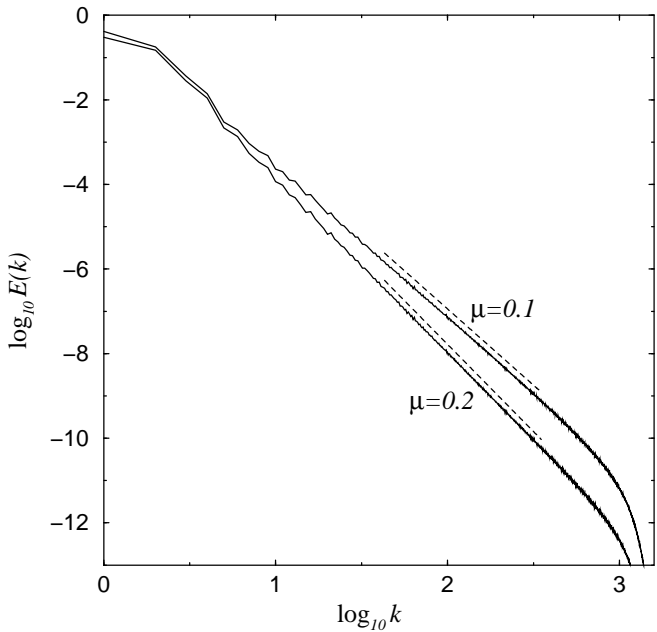}%
\caption{Energy wavenumber spectra. The dotted lines are the corresponding linear fit in
the scaling range.}
\label{fig:ek}
\end{figure}
different values of $\mu(k>6)$ we considered.
In both cases, a clear scaling range of more than a decade can be observed. We measure
the scaling exponents by linearly fitting $E(k)$ in the scaling range. The results, denoted
$\xi_{DNS}$, are shown in \T{tbl:xi}. Good agreement is found between the numerical and
the theoretical results. These results are consistent with those of previous work in \cite{nam00a}
and \cite{boffetta02} which use grids of $1024\times 1024$ and $2048\times 2048$, respectively.

\subsubsection{Vorticity Structure Functions}
\label{sec:s2q_com}

Numerical tests of theoretical predictions of structure function exponents of finite-lifetime
passive scalar fields advected by simple chaotic flows have been performed in Ref.~\cite{neufeld00}
and \cite{boffetta02}. To test the analogous theoretical predictions, \E{zeta2q} and \E{hq}, in
the case of two-dimensional turbulence with drag, we define the averaged structure functions of
order $2q$ as
\begin{equation}
\label{avgs2q}
S_{2q}(r)=\int\frac{d\V{r}\,'}{2\pi r}\delta(r-|\V{r}\,'|)
\langle\,|\delta_{\V{r}'}\omega|^{2q}\rangle\ .
\end{equation}
The angled brackets denote average over the entire region occupied by the fluid. The angular
dependence of $\langle\,|\delta_{\V{r}'}\omega|^{2q}\rangle$ is averaged out in \E{avgs2q} by
the integration over $\V{r}\,'$. Using \E{avgs2q} with $\omega(\V{x},t)$ obtained from the
numerical integration of \E{weq}, we compute $S_{2q}(r)$ from $t=41$ to $t=75$ at every 1 time
unit and take the average of the results obtained.

\begin{figure}
\includegraphics[scale=0.4]{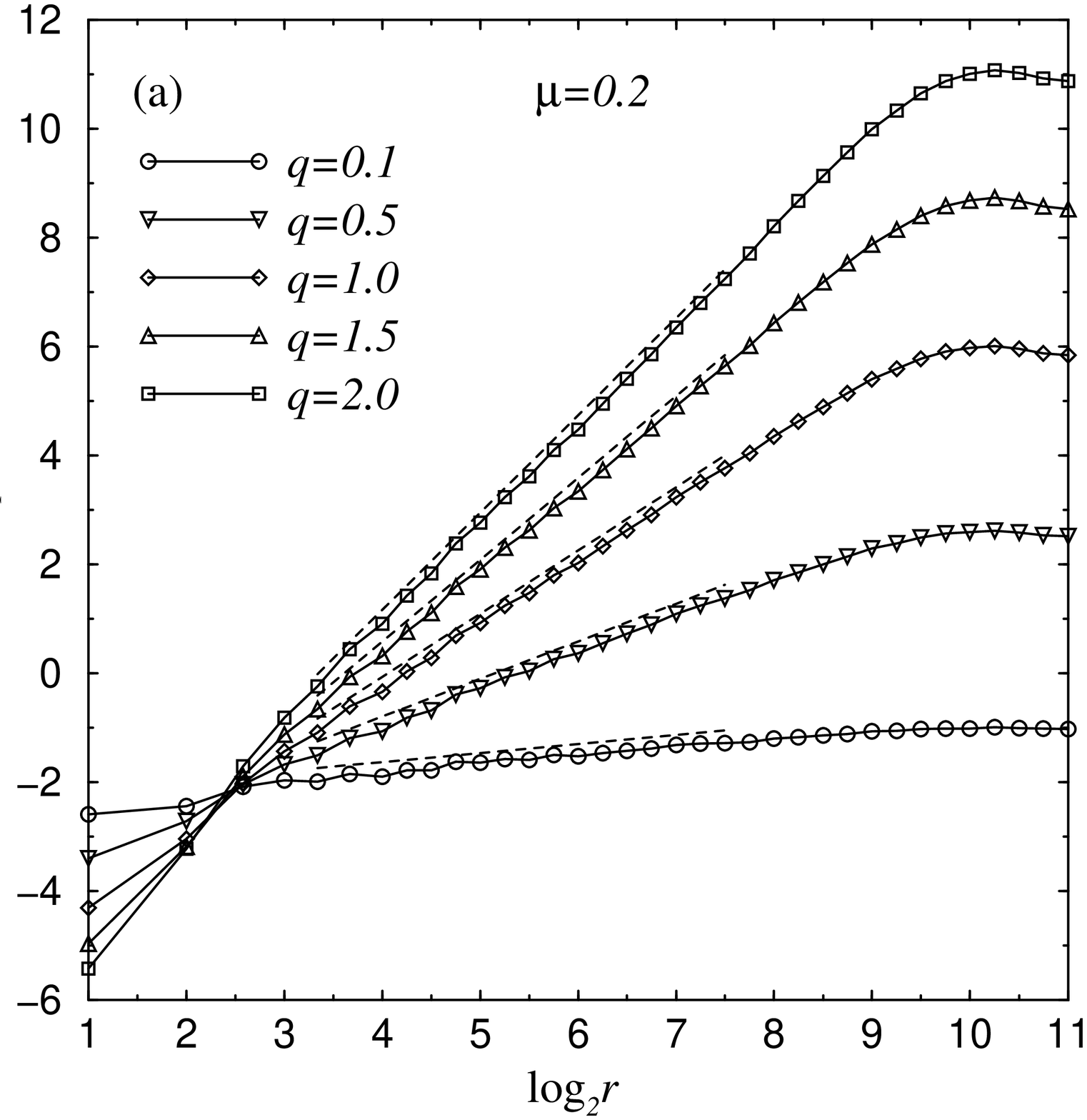}\\%
\vspace{0.6cm}
\includegraphics[scale=0.4]{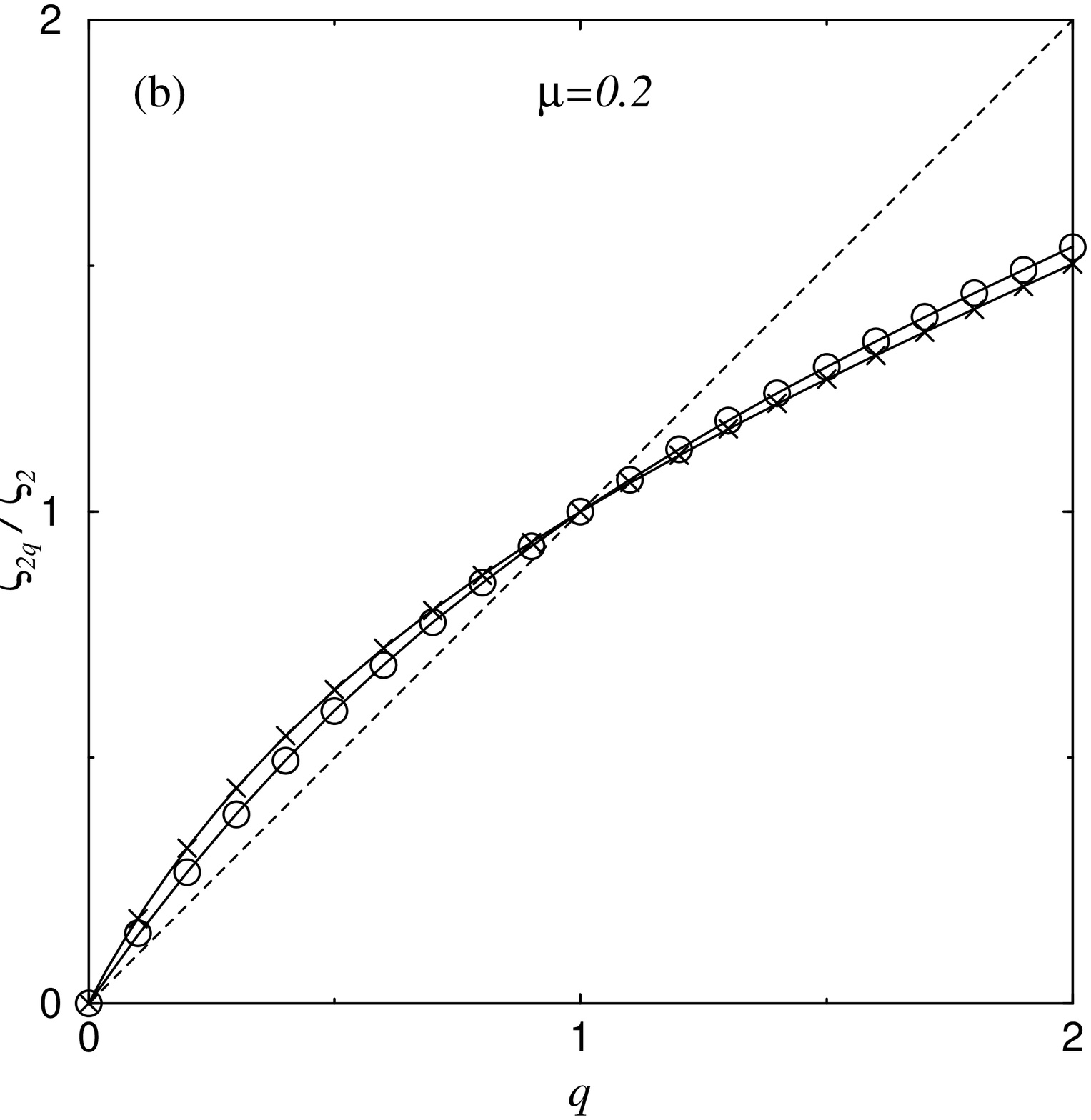}%
\caption{For the case of $\mu=0.2$: (a) structure functions of vorticity difference, for
various orders $q$ between 0.1 and 2.0; the dotted lines are linear fits in the scaling
range. (b) Plot of $\zeta_{2q}/\zeta_2$ obtained from numerical simulations (circle) and
from \E{zeta2q} (cross), for different values of $q$; the solid lines are
polynomial fits to the circles and the crosses (cf. \E{z2qpoly}).}
\label{s2}
\end{figure}

Following the scheme described above, we calculate the time-averaged structure functions
for $q$ ranging from 0.0 to 2.0. \FS{s1}(a) and \ref{s2}(a) show samples of the results
for the cases $\mu=$0.1 and 0.2, respectively. The distance $r$ is measured in the unit of
grid size.%
\begin{figure}
\includegraphics[scale=0.4]{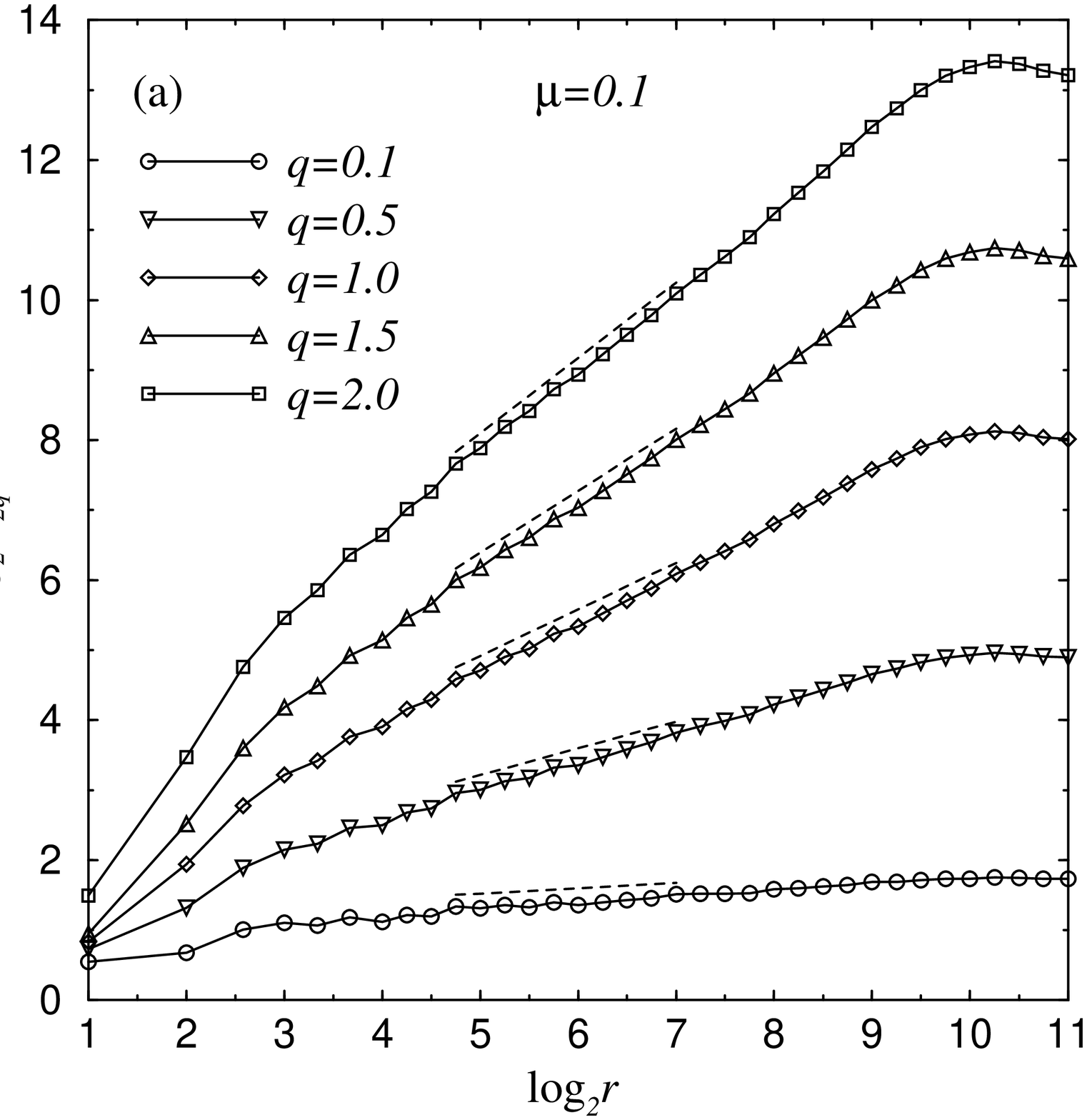}\\%
\vspace{0.6cm}
\includegraphics[scale=0.4]{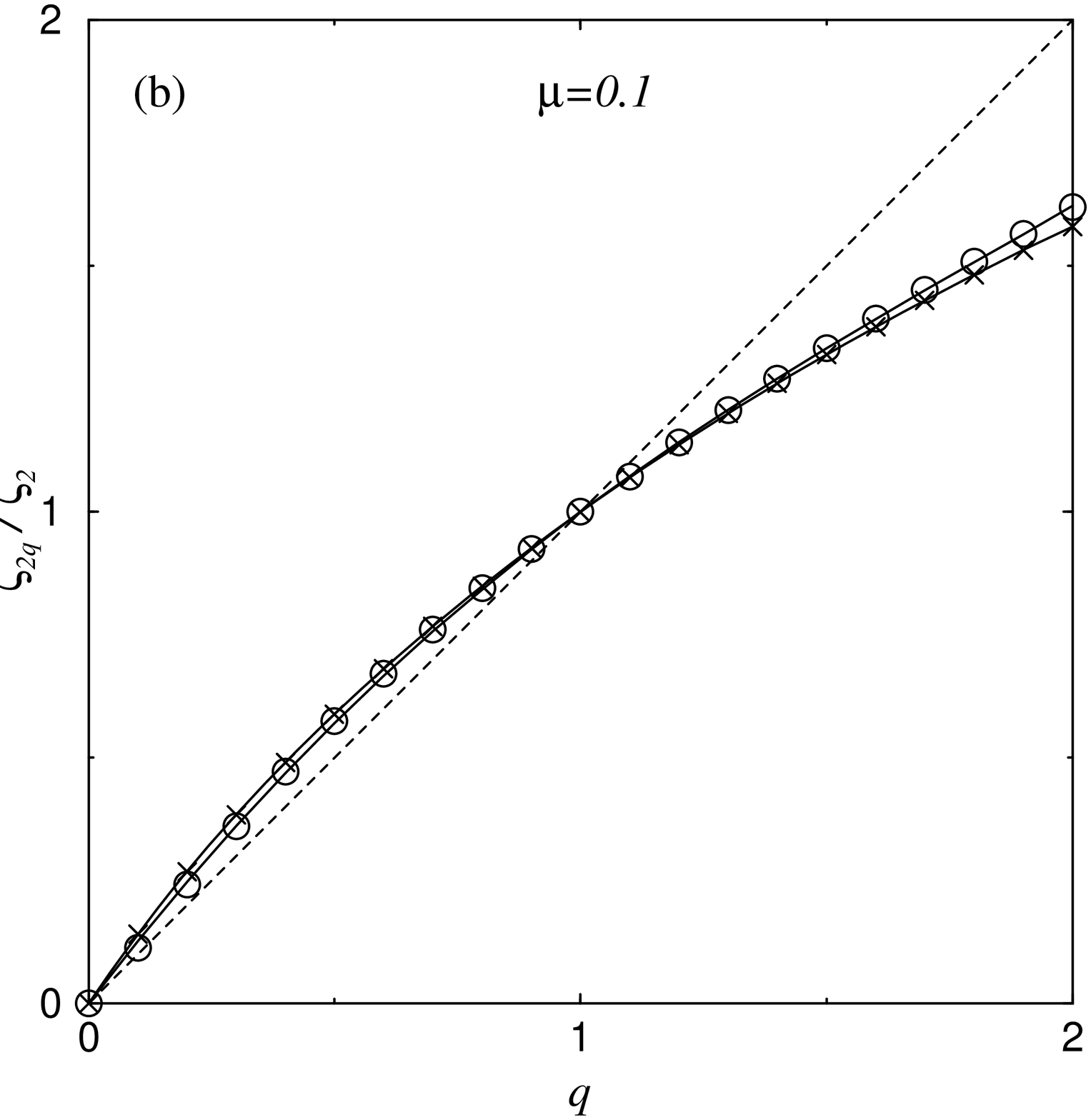}%
\caption{For the case of $\mu=0.1$: (a) structure functions of vorticity difference, for
various orders $q$ between 0.1 and 2.0; the dotted lines are linear fits in the scaling
range. (b) Plot of $\zeta_{2q}/\zeta_2$ obtained from numerical simulations (circle) and
from \E{zeta2q} (cross), for different values of $q$; the solid lines are
polynomial fits to the circles and the crosses (cf. \E{z2qpoly}).}
\label{s1}
\end{figure}
For all values of $q$ we have studied, the structure functions show a clear scaling
range that is long enough to allow an estimate of the scaling exponents, $\zeta_{2q}$. The
scaling range of the structure functions in real space roughly corresponds to that of the
energy spectrum in $k$-space. The values of $\zeta_{2q}$ are obtained by measuring the
slope of the structure functions in the scaling range using a linear fit. Results for
$\zeta_{2q}/\zeta_2$ are shown as circles in \FS{s1}(b) and \ref{s2}(b). The measured values
of $\zeta_2$, denoted as $\zeta_{2,DNS}$, are given in \T{tbl:xi}. We then obtain theoretical
predictions to $\zeta_{2q}$ from the polynomial fit of $\bar\zeta_{2q}$ shown in \F{z2q_ehzt},
following procedures described at the beginning of this section. The results are shown as crosses
in \F{s1}(b) and \ref{s2}(b) for the two cases of $\mu$ we studied. The predicted values of $\zeta_2$,
denoted as $\zeta_{2,th}$, are given in \T{tbl:xi}. The numerical and theoretical results agree
reasonably well for all $q$'s. In reference to \E{zeta2q} we note that, for all the cases we
numerically tested, we found that $\zeta_{2q}=\bar\zeta_{2q}<2q$ for all $q$. This is consistent
with the general result that for $\mu<\bar{h}$, $\bar\zeta_{2q}<H_q(\bar{h})<2q$. We note, however,
that this need not hold in general, particularly for large $\mu$.

\subsection{Discussion}

\subsubsection{Energy Wavenumber Spectrum}
\label{sec:ekdis}

The presence of drag gives a power law for the energy spectrum that falls faster than the case
without drag and $\xi$ is the correction to the classical zero drag value of the scaling exponent.
Note that the major contribution to the integrals involved in the theory \cite{nam00b}
comes from the immediate neighborhood of $h=h_1^*$, where $h_1^*$ is the value of $h$ at which
the minimum of $H_1(h)$ occurs. As $\mu\rightarrow 0$, $h_1^*\rightarrow\bar{h}$ and hence
$\xi\rightarrow 0$. The scaling exponent is thus determined by the majority of fluid trajectories
with stretching rate close to $\bar{h}$. On the other hand, for $\mu\neq 0$, $h_1^*>\bar{h}$.
Therefore the scaling exponent is determined by a small number of fluid trajectories that
experience a stretching rate $h_1^*$, that is larger than the typical stretching rate $\bar{h}$,
and $h_1^*$ increases with $\mu$ ($h_1^*$ for $\mu=0.1$ and $\mu=0.2$ are 0.47 and 0.57 respectively).
The reason for this, and in general $h^*_q>\bar h$ as we have shown in Section~\ref{sec:ek_sq_th},
is that the presence of drag introduces the exponential decaying factor $e^{-\mu\tau}$ in \E{avgt} and
the corresponding integral for $E(k)$ \cite{nam00b}. Thus, for a certain fixed $k$ (or $\lambda$), the
fluid trajectories that contribute most to $E(k)$ (or $S_{2q}(r)$) are those with smaller $\mu\tau$,
and hence larger stretching rate $h$.

It should also be clear from the above discussion that, in order to get an accurate estimate to $\xi$,
it is important to take into account the fluctuation in the finite-time Lyapunov exponent. If we were
to neglect the fluctuation and take the stretching rate to be $\bar{h}$ for all trajectories, we would
have overestimate $\xi$ to be $2\mu/\bar{h}$.

\subsubsection{Vorticity Structure Functions}
\label{sec:s2qdis}

The theory predicts that when drag is present, an inertial range exists in
which the vorticity structure functions $S_{2q}(r)$ exhibit power-law scaling. The scaling
exponent $\zeta_{2q}$ is given by \E{zeta2q}. In the absence of intermittency $\zeta_{2q}$
scales linearly with $q$ and $\zeta_{2q}/\zeta_2=q$, which is plotted as the straight dashed
line in \FS{s1}(b) and \ref{s2}(b). For $\mu > 0$, \E{zeta2q} predicts
that $\zeta_{2q}$ varies nonlinearly with $q$, as shown in \FS{s1}(b)
and \ref{s2}(b). This anomalous scaling of $S_{2q}(r)$, which indicates the presence of
intermittency in the system, is verified numerically in both cases.

The anomalous scaling of $S_{2q}(r)$ is the result of the combined effect of drag and
non-uniform stretching of fluid elements. If all fluid elements have the same stretching rate,
say $\bar{h}$, then $e^{-\mu\tau}$ becomes a constant and thus $\zeta_{2q}$ is proportional
to $2q$. On the other hand, if $\mu=0$, regardless of whether the stretching is uniform or not,
$\zeta_{2q}=0$. \FS{s1}(b) and \ref{s2}(b) also suggest that $S_{2q}(r)$ shows larger deviation
from the simple scaling behavior, $\zeta_{2q}/\zeta_2=q$, as $\mu$ increases.

We remark that the statistical error in the predictions of the values of the higher order
structure function scaling exponents is in general larger. This is because, as already mentioned
in Section~\ref{sec:ek_sq_th}, $h_q^*$ increases with $q$, hence for large $q$, $\zeta_{2q}$ depends
mostly on a rare number of fluid trajectories with large $h$.

\subsubsection{$S_2(r)$ and $E(k)$}
\label{sec:s2ek}

We now focus on the second order structure function $S_2(r)$. With the isotropic assumption,
the following relation between $S_2(r)$ and $E(k)$ can be obtained,
\begin{equation}
S_2(r)=4\int\! dk\,[1-J_0(kr)]k^2 E(k)
\label{s2e}
\end{equation}
where $J_0$ is the zeroth order Bessel function of the first kind. We compute $S_2(r)$ by \E{s2e}
using the $E(k)$ shown in \F{fig:ek}. The results are plotted as solid lines in \F{s2_e},%
\begin{figure}
\includegraphics[scale=0.4]{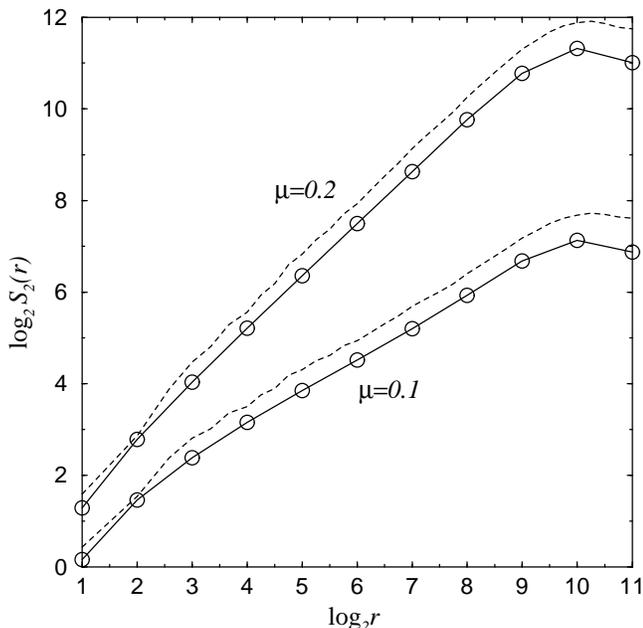}%
\caption{Second order structure functions $S_2(r)$ obtained from the energy spectrum $E(k)$
using \E{s2e} (solid line with circles) and from direct numerical simulations (dotted line).}
\label{s2_e}
\end{figure}
the scaling exponents, denoted as $\zeta_{2,E(k)}$, are then measured and compared with
$\zeta_{2,DNS}$ in \T{tbl:xi}. The $S_2(r)$ obtained directly from \E{avgs2q} are plotted
as dashed lines in the same diagram for comparison. Ignoring viscosity at
the small scales and forcing at the large scales, we can assume $E(k)\sim k^{-(3+\xi)}$ for
all $k$. Then it follows from \E{s2e} that, for $0<\xi<2$, $S_2(r)\sim r^{\zeta_2}$ for all
$r$ and $\zeta_2=\xi$. This is consistent with our theory which predicts that
$\zeta_{2,th}=\xi_{th}$ when $0<\zeta_{2,th}<2$. Comparing $\zeta_{2,DNS}$ to $\xi_{DNS}$,
we find reasonably good agreements but we note that $\zeta_{2,DNS}$ is in general slightly
larger than $\xi_{DNS}$. We shall see that this small discrepancy is a result of the dependence
of $S_2(r)$ on the large scales of the flow.

Due to the effects of forcing and viscosity, $E(k)$ deviates from the power-law behavior at
very small and very large $k$. According to \E{s2e}, these deviations will affect $S_2(r)$.
This is demonstrated in \F{s2ek} in which we plot the $S_2(r)$ calculated
by \E{s2e} using three different $E(k)$.%
\begin{figure}
\includegraphics[scale=0.4]{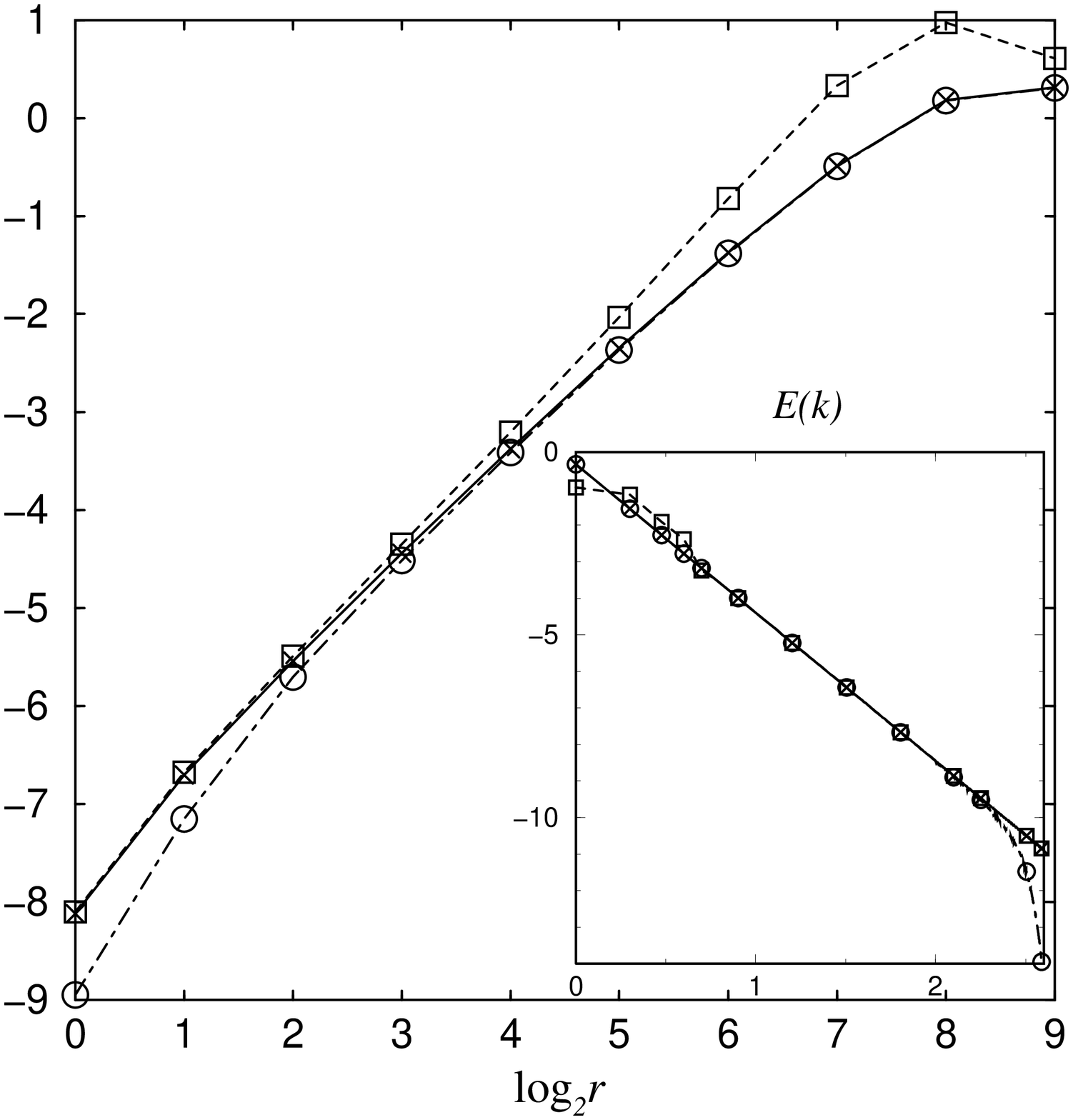}%
\caption{Second order structure functions $S_2(r)$ calculated by \E{s2e} using different
$E(k)$: pure power law (cross), power law with small scales viscous cutoff (circle) and
power law with large scales fluctuations (square). The inset shows the different $E(k)$
versus $k$ with corresponding labels.}
\label{s2ek}
\end{figure}
The first one is a pure power law for all $k$. The second
one is the same as the first one except it drops faster at large $k$ mimicking the viscous
cutoff. The third one is constructed from the first one by modifying the first five modes
to mimic the presence of large scale forcing. The range of $k$ used corresponds to a
$1024\times 1024$ lattice. From \F{s2ek}, it is seen that while the viscous effect is
negligible except at small $r$, the large scale forcing can have a significant effect on the
general shape of $S_2(r)$. As a result, $S_2(r)$ may have a very limited scaling range even
when $E(k)$ shows a reasonably long one. We believe this is generally true for structure
functions of any order, making it more difficult to accurately measure $\zeta_{2q}$ than $\xi$.
We are able to obtain reliable estimates of $\zeta_{2q}$, as verified by the agreement between
$\zeta_{2,DNS}$ and $\xi_{DNS}$, by employing a $4096\times 4096$ lattice and the time-averaging
process.

\section{Probability Distribution of Vorticity Difference}
\label{sec:pdf}
\subsection{Theory}

In this section, we shall derive an expression for the probability distribution function
$P_r(\delta_{\V{r}}\omega)$ of the vorticity difference $\delta_{\V{r}}\omega$ in terms of
the conditional probability density function $R(\tau\,|\,\lambda)$. In Section~\ref{sec:ek_sq_th},
we introduced $R(\tau\,|\,\lambda)$ and deduced the scaling of $\delta_{\V{r}}\omega$ by doing order
of magnitude estimations of the integrals in \E{deltaw}. We now determine the distribution
of $\delta_{\V{r}}\omega$ by estimating the values of these integrals for different fluid
trajectories. We shall assume the system is isotropic so that $P_r(\delta_{\V{r}}\omega)$
depends on the distance $r$ only, and not on the direction of $\V{r}$.

Referring to \E{deltaw}, for $t-\tau<t'<t$, we estimate that $\V{r}\cdot\nabla f_\omega[\V{x}(t')]
\sim(r/L)f_\omega[\V{x}(t')]=e^{-\lambda}f_\omega[\V{x}(t')]$. For $0<t'<t-\tau$, the separation
between the two trajectories is of order $L$ and the values of $f_\omega[\V{x}(t')]$ along the
two trajectories become uncorrelated, hence $\delta f_\omega[\V{x}(t')]\sim f_\omega[\V{x}(t')]$.
Approximating $f_\omega[\V{x}(t')]$ as varying randomly, the values of the first and the second
integral in \E{deltaw} are respectively estimated as $[f_\omega/(\mu-h)](e^{-\lambda}-e^{-\mu\tau})$
and $(f_\omega/\mu)e^{-\mu\tau}$. This suggests that we can treat $\delta_{\V{r}}\omega$
as a function of two random variables $\tau$ and $Z=f_\omega/\mu$ as follow,
\begin{equation}
\delta_{\V{r}}\omega(Z,\tau)\approx Z\,\frac{e^{-\lambda}+e^{-\mu\tau}}{2}\ .
\label{dwzt}
\end{equation}
We then write the probability distribution function of $\delta_{\V{r}}\omega$ as,
\begin{equation}
P_r(\delta_{\V{r}}\omega)=
\int_0^\infty P_r(\delta_{\V{r}}\omega|\tau)R(\tau\,|\,\lambda)\,d\tau
\end{equation}
where $P_r(\delta_{\V{r}}\omega|\tau)$ is the conditional probability distribution function of
$\delta_{\V{r}}\omega$ given $\tau$. From \E{dwzt} and letting $y(\tau)=(e^{-\lambda}+e^{-\mu\tau})/2$,
we have,
\begin{equation}
P_r(\delta_{\V{r}}\omega|\tau)
=\frac{1}{y(\tau)}P_Z\left[\frac{\delta_{\V{r}}\omega}{y(\tau)}\right]
\end{equation}
where $P_Z$ is the probability density function of $Z$. Note that when $r=L$,
$\lambda=\tau=0$, which implies $\delta_{\V{r}}\omega|_{r=L}=Z$. Therefore, we obtain the result,
\begin{equation}
P_r(\delta_{\V{r}}\omega)=\int_0^\infty
\frac{1}{y(\tau)}P_L\left[\frac{\delta_{\V{r}}\omega}{y(\tau)}\right]R(\tau\,|\,\lambda)\,d\tau\ .
\label{pdw_th}
\end{equation}
Reference \cite{chertkov98} has previously derived the probability distribution function for
differences of finite lifetime passive scalars advected by a temporally white noise model
velocity field. For the model used in Ref.~\cite{chertkov98} the difference probability distribution
function at large separation $|\vec{r}|$ is Gaussian, while this is not the case for our flow.
In order to obtain reasonable agreement between the theory \E{pdw_th} and our numerical experiments,
it is important to account for the non-Gaussian behavior of the probability distribution function
of the vorticity difference at large separation $|\vec{r}|$. In what follows, this will be done
by using the numerically obtained probability distribution function at large $|\vec{r}|$ as an
input to \E{pdw_th} to determine the probability distribution function at small $|\vec{r}|$.

\subsection{Comparison of Theory and Numerical Results}

We first compute the probability distribution function $\bar{P}_r(X_{\V{r}})$ of
the standardized vorticity difference $X_{\V{r}}$, \E{xr}, directly from the
numerical solution of \E{weq}. The vorticity field $\omega(\V{x},t)$ from $t=41$
to $t=75$ at every 1 time unit is used in this computation. The separating
distance $\V{r}$ is taken to be in the $\hat{x}$-direction and is measured in the
unit of grid size. For $\mu=0.1$, the results for four different values of $r$ are
shown as solid lines in \F{pdw1}.%
\begin{figure*}
\includegraphics[scale=0.7]{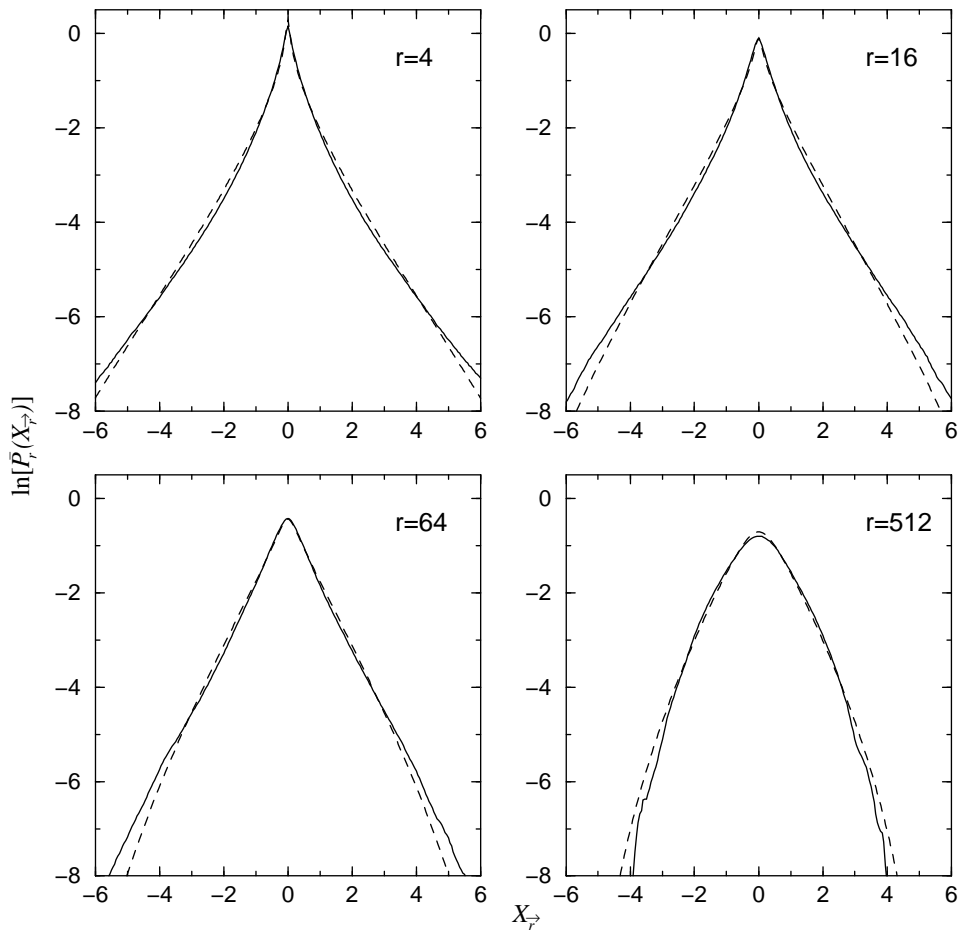}%
\caption{For $\mu=0.1$, the probability distribution function $\bar{P}_r(X_{\V{r}})$ of the standardized
vorticity difference $X_{\V{r}}$ obtained from direct numerical simulation (solid lines) and from
\E{pdw_th} (dashed lines). The separating distance $\V{r}$ is in the $\hat{x}$-direction and measured
in the unit of grid size.}
\label{pdw1}
\end{figure*}
It is clear that the shape of $\bar{P}_r(X_{\V{r}})$ changes as $r$ varies in the range
$4\leq r \leq 512$, indicating the system is intermittent. $\bar{P}_r(X_{\V{r}})$ develops
exponential tails (e.g., $r=64$) and then stretched-exponential tails (e.g., $r=4$) as $r$
decreases. We note that $\bar{P}_r(X_{\V{r}})$ for all $r \leq 4$ collapse onto each other,
and similarly, all the $\bar{P}_r(X_{\V{r}})$ with $r \geq 512$ have the same shape. Numerical
results similar to those in \F{pdw1} have also been obtained in Ref.~\cite{boffetta02}, but
theory for $\bar{P}_r(X_{\V{r}})$ was not presented in Ref.~\cite{boffetta02}.

To apply the theoretical result \E{pdw_th}, we first derive an expression for $R(\tau\,|\,\lambda)$.
To this end, we approximate $G(h)$ as a quadratic function of $h$,
\begin{equation}
G(h) \approx a(h-\bar{h})^2\ .
\label{gh2}
\end{equation}
Using the relation \E{rtlapp} and the asymptotic form for $P(h\,|\,t)$, \E{gh}, we obtain
\begin{eqnarray}
R(\tau\,|\,\lambda)&=&\frac{\lambda}{2\tau^2}P(h=\frac{\lambda}{\tau}\,|\,\tau) \nonumber \\*
&\sim&\frac{1}{2}\sqrt{\frac{a}{\pi\tau}}\left(\frac{\lambda}{\tau}-\bar{h}\right)
e^{-\tau a\left(\frac{\lambda}{\tau}-\bar{h}\right)^2}\ .
\label{rtl2}
\end{eqnarray}
To compare the predictions by \E{pdw_th} with numerical results, the $G(h)$ for $\mu=0.1$ is
obtained numerically as described in Section~\ref{sec:pht} and fitted to \E{gh2} in the vicinity of
its minimum to obtain the parameters $a$ and $\bar{h}$. This quadratic fit is a good approximation
because the integral in \E{pdw_th} is dominated by the region where $R(\tau\,|\,\lambda)$ is large,
which roughly corresponds to the region where $P(h\,|\,t)$ is maximum. \F{rtl1} shows the
$R(\tau\,|\,\lambda)$ for $\mu=0.1$ obtained using
\begin{figure}
\includegraphics[scale=0.4]{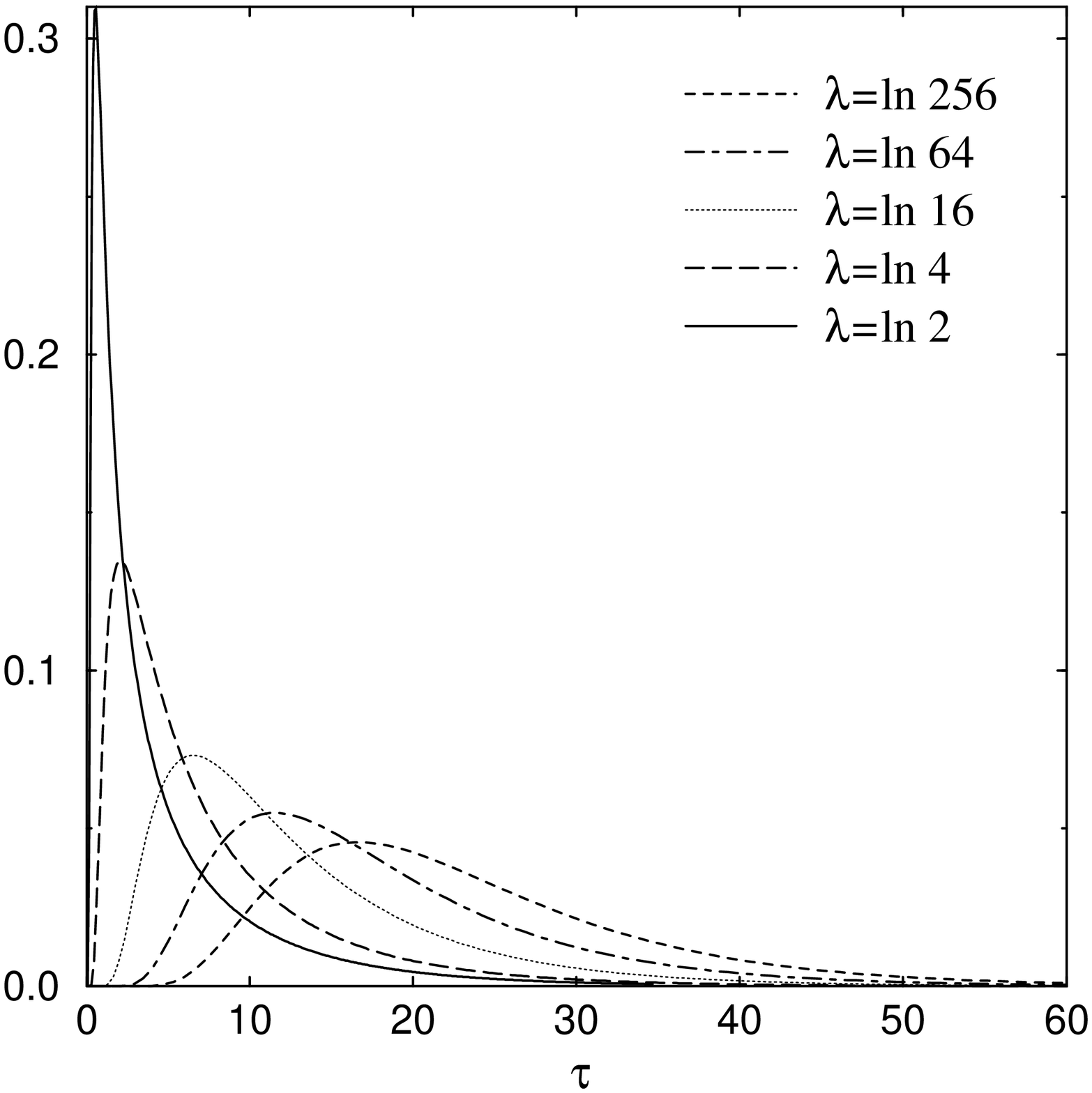}%
\caption{For $\mu=0.1$, the conditional probability density function $R(\tau\,|\,\lambda)$ given
by \E{rtl2} ($a=1.55$, $\bar{h}=0.26$) for different values of $\lambda$.}
\label{rtl1}
\end{figure}
the quadratic approximation \E{gh2}. We then take $P_L$ in \E{pdw_th} to be of the
form $\exp(W)$ where $W$ is an even sixth degree polynomial fitted to the
numerically computed $\ln[P_r(\delta_{\V{r}}\omega)]$ for $\V{r}=1024\hat{x}$ and
$\mu=0.1$. With $R(\tau\,|\,\lambda)$ and $P_L$ obtained as described above, we
compute $P_r(\delta_{\V{r}}\omega)$, and thus $\bar{P}_r(X_{\V{r}})$, for
different values of $r$ using \E{pdw_th}. The results are plotted as dashed lines
in \F{pdw1}. The theoretical predictions agree well with the numerical results. We
also find good agreements between our theory and numerical simulations when
$\V{r}$ is taken to be in the $\hat{y}$-direction. Similar results were also
obtained for the case $\mu(k>6)=0.2$.

\subsection{Discussion}

According to \E{pdw_th}, $P_r(\delta_{\V{r}}\omega)$ for a given $r$ can be considered as a
superposition of many different probability distribution functions, each obtained by rescaling 
$P_L(\delta_{\V{r}}\omega)$ to a different width using $y(\tau)$. The amplitude of each component
is then determined by the coefficient $R(\tau\,|\,\lambda)$. Since the statistics between two
points separated by distance $r \sim L$ are uncorrelated, the distribution $P_L(\delta_{\V{r}}\omega)$
is virtually the one-point probability distribution of the vorticity $\omega$, which is not necessarily
Gaussian and is closely related to the statistics of the source $f_\omega$, as implied by the definition
of $Z$. Hence, \E{pdw_th} relates the distribution of the vorticity difference to the one-point
statistics of the source.

As can be seen in \F{rtl1}, $R(\tau\,|\,\lambda)$ has a very sharp peak when $\lambda$ is small
($r$ is large). Thus, only a small range of values of $\tau$ contributes to the integral \E{pdw_th}.
This gives the expected result that $P_r(\delta_{\V{r}}\omega)$ has similar shape as
$P_L(\delta_{\V{r}}\omega)$ when $r$ is large. As $\lambda$ increases ($r$ decreases), the width
of $R(\tau\,|\,\lambda)$ increases. According to \E{pdw_th}, $P_r(\delta_{\V{r}}\omega)$
is now composed of a large number of rescaled $P_L(\delta_{\V{r}}\omega)$ with a broad range of width.
This gives rise to the broader-than-Gaussian tails in
$P_r(\delta_{\V{r}}\omega)$ for small $r$.

Like the anomalous scaling of the vorticity structure functions, the dependence of the shape
of $P_r(\delta_{\V{r}}\omega)$ on $r$ is a result of the presence of drag and the non-uniform
stretching of fluid elements. If $h=\bar{h}$ for all fluid trajectories, then
$R(\tau\,|\,\lambda)=\delta(\tau-\lambda/\bar{h})$, and from \E{pdw_th}, $P_r(\delta_{\V{r}}\omega)$
has the same form for all $r$, indicating a self-similar flow. On the other hand, if $\mu=0$,
$y(\tau)$ becomes independent of $\tau$, which also implies $P_r(\delta_{\V{r}}\omega)$ is
independent of $r$. In both situations, $P_r(\delta_{\V{r}}\omega)$ will have the same shape as
$P_L(\delta_{\V{r}}\omega)$.

We remark that while the width of the distribution $P_r(\delta_{\V{r}}\omega)$ in \E{pdw_th}
depends precisely on the choice of the value for the multiplicative constant on the right hand side
of \E{dwzt}, our results for $\bar{P}_r(X_{\V{r}})$ are independent of this value.

\section{Multifractal Formulation}
\label{sec:dq}
\subsection{Theory}

The local viscous energy dissipation rate per unit mass $\varepsilon$ and its global average
value $\avg{\varepsilon}$ play important roles in the phenomenology of three-dimensional
turbulence \cite{stolovitzky94}. It is now well known that $\varepsilon$ shows intermittent
spatial fluctuations which can be described by the multifractal formulation \cite{frisch95}.
Using the measure $p'(r)=\mathcal{E}_r/\mathcal{E}$, where $\mathcal{E}_r$ is the total dissipation
in a volume of linear dimension $r$ and $\mathcal{E}$ is the total dissipation in the whole domain,
the R\'{e}nyi dimension spectrum $D_q$ and the singularity spectrum $f(\alpha)$ have been
measured experimentally \cite{chhabra89b,meneveau91}.

Intermittency in three-dimensional turbulence also manifests itself as anomalous scaling in
the velocity structure functions $S'_{3q}(r)$ defined as
\begin{equation}
S'_{3q}(r)=\avg{(\delta_{\V{r}}v)^{3q}} \sim r^{\zeta'_{3q}}\ ,
\end{equation}
where $v$ is the component of the velocity vector in the direction of $\V{r}$ and
$\delta_{\V{r}}v=v(\V{x}+\V{r})-v(\V{x})$. From Kolmogorov's hypotheses in his 1941 paper
\cite{k41}, which ignore the intermittency of $\varepsilon$, one arrives at the result $\zeta'_{3q}=q$.
Experiments show that $\zeta'_{3q}$ is a nonlinear function of $q$. This anomalous scaling of
$S'_{3q}(r)$ is believed to be related to the intermittency of $\varepsilon$. Kolmogorov's `refined
similarity hypothesis' in his 1962 paper \cite{k62} gives the connection between intermittency in
velocity difference and intermittency in $\varepsilon$. The refined similarity hypothesis states
that at very high Reynolds numbers, there is an inertial range of $r$ in which the conditional
moments of $\delta_{\V{r}}v$ scales as follows
\begin{equation}
\avg{(\delta_{\V{r}}v)^q\ \vline\ \varepsilon_r} \sim (r\varepsilon_r)^{q/3}\ ,
\label{rsh}
\end{equation}
where $\varepsilon_r$ is the average of $\varepsilon$ over a volume of linear dimension $r$.
\E{rsh} implies $\avg{(\delta_{\V{r}}v)^q} \sim \avg{(r\varepsilon_r)^{q/3}}$ which gives the
following relation between the $D_q$ based on $p'(r)$ and $\zeta'_{3q}$,
\begin{equation}
D_q=3+\frac{\zeta'_{3q}-q\zeta'_3}{q-1}\ .
\label{dqf-3d}
\end{equation}
Kolmogorov's fourth-fifths law gives $\zeta'_3=1$ exactly \cite{frisch95}. \E{dqf-3d} can also
be derived \cite{meneveau91} from the relation
\begin{equation}
\varepsilon_r \sim \frac{(\delta_{\V{r}}v)^3}{r}\ .
\label{er}
\end{equation}
We note that \E{rsh} follows from \E{er}. As shown below, there is a relation analogous to \E{dqf-3d}
for the enstrophy cascade of two-dimensional turbulence with drag.

For the enstrophy cascade regime in two-dimensional turbulence, a central quantity to the
phenomenology \cite{kraichnan67} in this regime is the viscous enstrophy dissipation $\eta$ given
by \E{eta}, and the relevant measure is $p_i(\epsilon)$ defined in \E{pidef}. We have already seen
in Section~\ref{sec:xi_zeta} that in the presence of drag, the vorticity structure functions scale
anomalously with scaling exponents $\zeta_{2q}$ given by \E{zeta2q}. We now derive a relation
between $\zeta_{2q}$ and the $D_q$ based on $p_i(\epsilon)$.

Consider the following quantity
\begin{equation}
I_1(q,\epsilon)=\sum_i p_i^q(\epsilon)
\label{iqedef}
\end{equation}
which by the definition of $D_q$, \E{renyidq}, scales like
\begin{equation}
I_1(q,\epsilon) \sim \epsilon^{(q-1)D_q}
\label{iqe}
\end{equation}
for some range of $\epsilon$. Assume there exists a scaling range extending from the system
scale $L\sim k_f^{-1}$ down to the dissipative scale $r_d\sim k_d^{-1}$ such that both the scaling
relations \E{sqdef} and \E{iqe} hold. At the dissipative scale, due to the action of viscosity,
the vorticity field $\omega$ becomes smooth, thus we have the following relations,
\begin{align}
\displaystyle\int_{\mathcal{R}(r_d)}\!\!|\nabla\omega|^2d\V{x}\ &\sim\ r_d^2\,|\nabla\omega(\V{x})|^2\ ,
&& \V{x} \in \mathcal{R}(r_d) \\
|\nabla\omega|\ &\sim\ \displaystyle\frac{|\delta_{\V{r}}\omega|}{r_d}\ , && |\V{r}|=r_d\ .
\end{align}
Then by putting $\epsilon=r_d$ in \E{pidef} and let $|\V{r}|=r_d$ and $\V{x}_i\in \mathcal{R}_i(r_d)$,
we get
\begin{equation}
p_i(r_d) \sim
\frac{|\delta_{\V{r}}\omega(\V{x}_i)|^2}{(r_d)^{-2}\int_\mathcal{R}|\delta_{\V{r}}\omega|^2\,d\V{x}}\ .
\label{piapp}
\end{equation}
Substituting \E{piapp} in \E{iqedef}, we obtain the scaling of $I_1(q,\epsilon)$,
\begin{eqnarray}
I_1(q,r_d) &\sim& \frac{\displaystyle\sum_i|\delta_{\V{r}}\omega(\V{x}_i)|^{2q}}
{\displaystyle\frac{1}{(r_d)^{2q}}
\left(\displaystyle\int_\mathcal{R}|\delta_{\V{r}}\omega|^2\,d\V{x}\right)^q}
\nonumber \\*
\nonumber\\*
&\sim& \frac{\displaystyle\left(\frac{L}{r_d}\right)^2\avg{|\delta_{\V{r}}\omega|^{2q}}}
{\displaystyle\left(\frac{L}{r_d}\right)^{2q}\avg{|\delta_{\V{r}}\omega|^2}^q} \nonumber \\*
\nonumber\\*
&\sim& (r_d)^{2q-2+\zeta_{2q}-q\zeta_2}\ .
\label{iscale}
\end{eqnarray}
Comparing \E{iqe} to \E{iscale}, we get the principal result of this section,
\begin{equation}
D_q=2+\frac{\zeta_{2q}-q\zeta_2}{q-1}
\label{dqf-2d}
\end{equation}
which can be regarded as analogous to \E{dqf-3d}.

We mention that \E{dqf-3d} can be derived in an analogous manner using \E{er}.
From \E{dqf-2d}, in two-dimensional turbulence, the measure $p_i$ is multifractal 
when the vorticity structure functions exhibit anomalous scaling. Hence, in the presence of
drag, we expect the measure based on the squared vorticity gradient $|\nabla\omega|^2$ to show
multifractal structures.

\subsection{Comparison of Theory and Numerical Results}

The multifractal structure of $|\nabla\omega|^2$ is most readily visualized in snapshots of
$|\nabla\omega|^2$ from our simulations. Since $|\nabla\omega|^2$ grows at widely varying
exponential rates, only a few points would be visible if $|\nabla\omega|^2$ was plotted
directly using a linear scale. Therefore, we plot the following quantity instead \cite{varosi91},
\begin{equation}
\mathcal{M}(\V{x})=
\frac{\sum_{\V{x}_i\in \Lambda_{\V{x}}}|\nabla\omega(\V{x}_i)|^2}{\sum_{i}|\nabla\omega(\V{x}_i)|^2}\ ,
\end{equation}
where the set $\Lambda_{\V{x}}$ contains those lattice points $\V{x}_i$ for which
$|\nabla\omega(\V{x}_i)|^2 \leq |\nabla\omega(\V{x})|^2$, and we sum over all lattice points in
the denominator. By definition, $0\leq\mathcal{M}(\V{x})\leq 1$. \F{dw} shows the results for
$\mu=0.1$ and $\mu=0.2$.%
\begin{figure}
\includegraphics[scale=0.415]{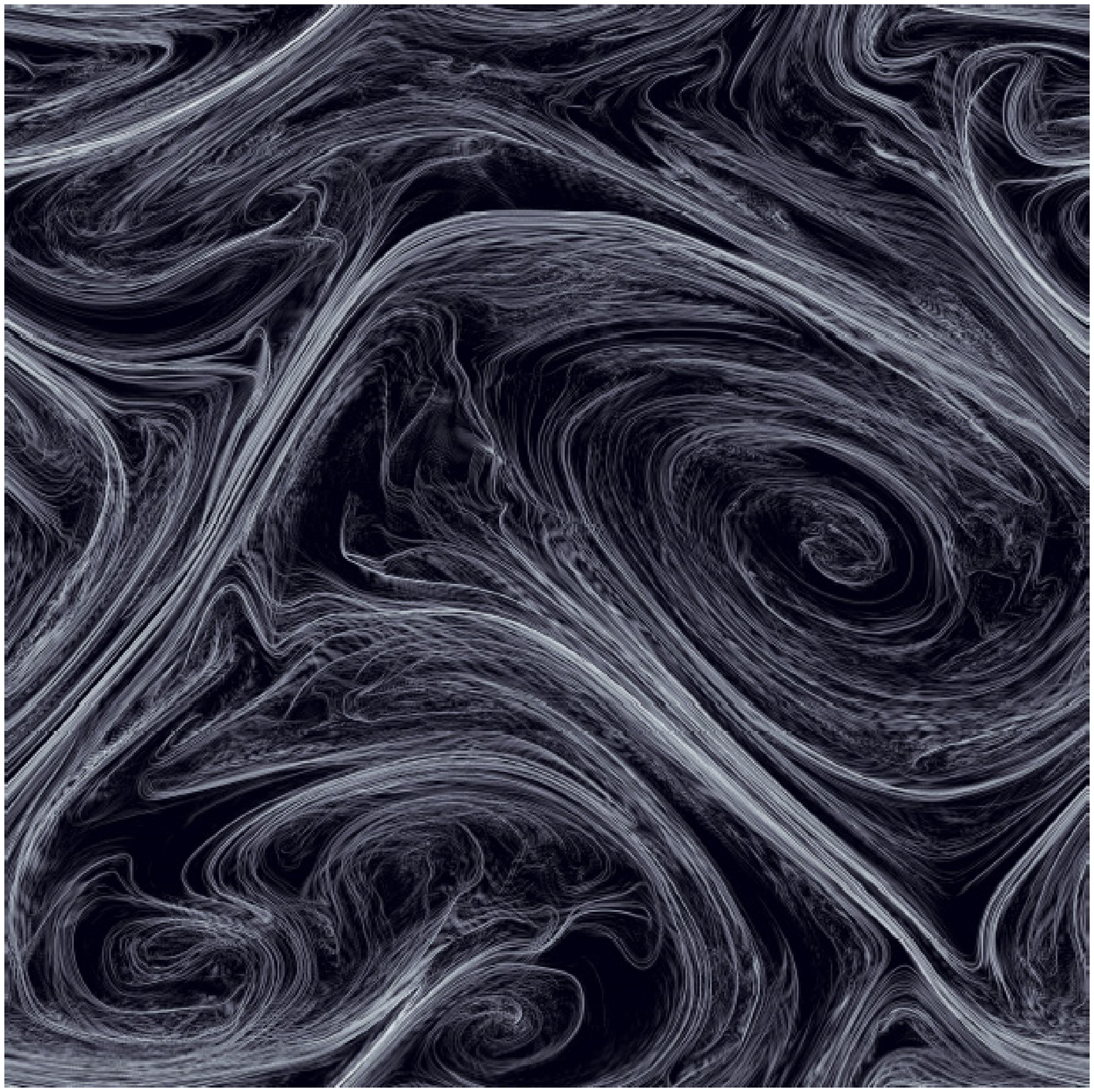}\\%
\vspace{0.3cm}
\includegraphics[scale=0.415]{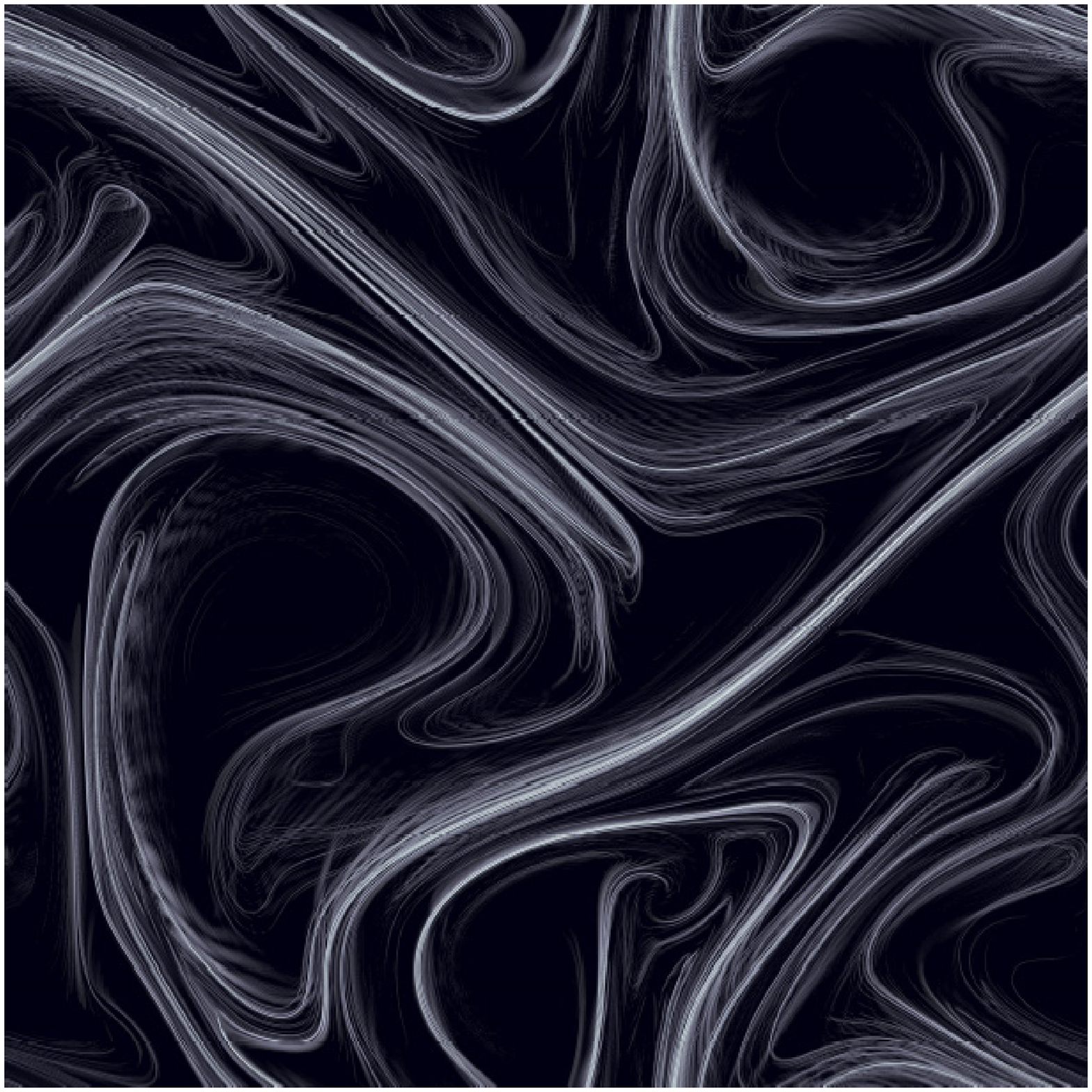}%
\caption{Snapshots of the scaled squared vorticity gradient $|\nabla\omega|^2$ at $t=61$ for
the case $\mu=0.1$ (upper) and at $t=65$ for the case $\mu=0.2$ (lower). Light areas are regions
of large values, and dark areas are regions of small values.}
\label{dw}
\end{figure}
Filament structures can clearly be seen for both cases, showing that the measure $p_i$ concentrates
in a very small area. This is particularly clear in the case $\mu=0.2$.

\begin{figure}
\includegraphics[scale=0.4]{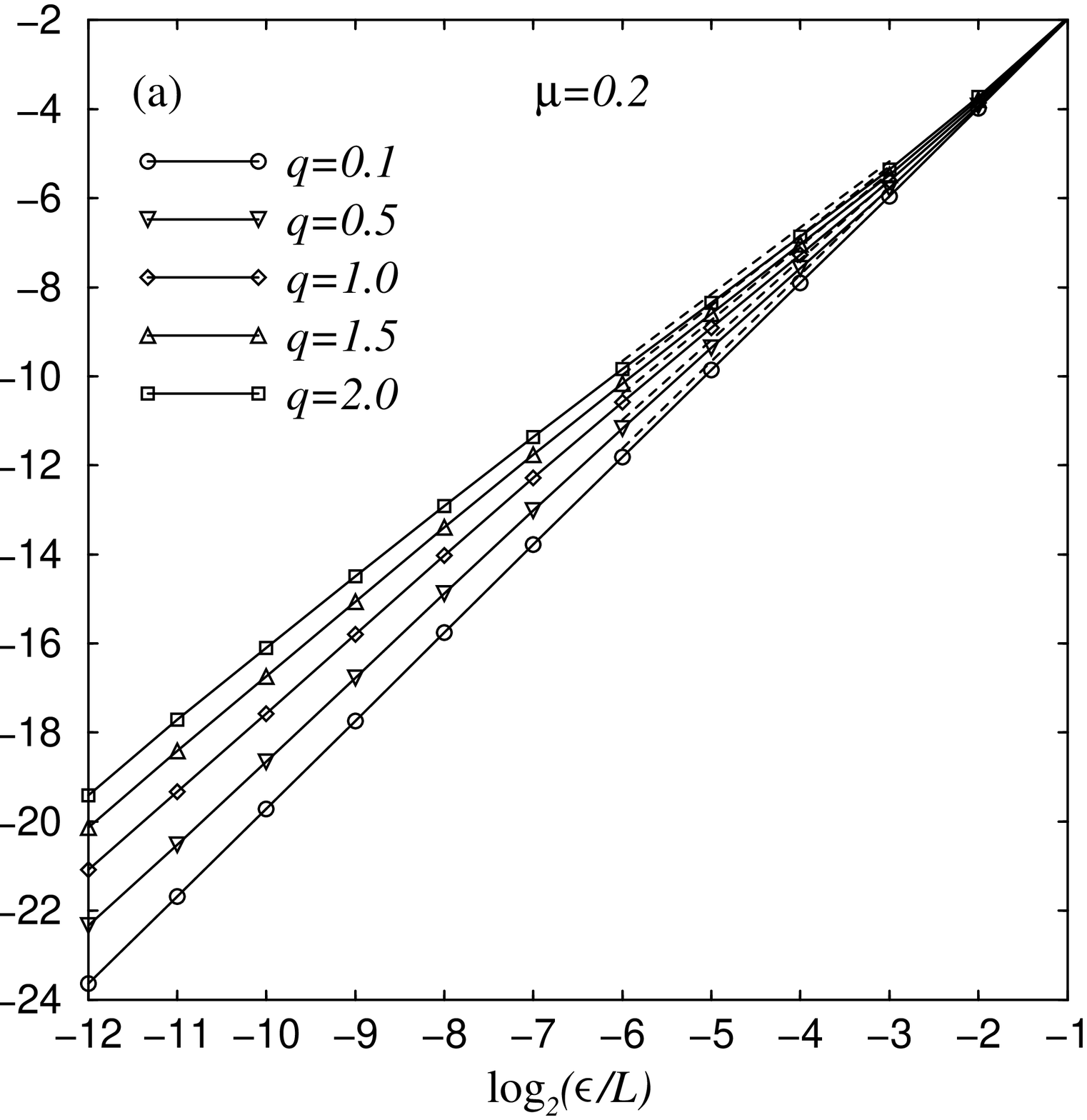}\\%
\vspace{0.6cm}
\includegraphics[scale=0.4]{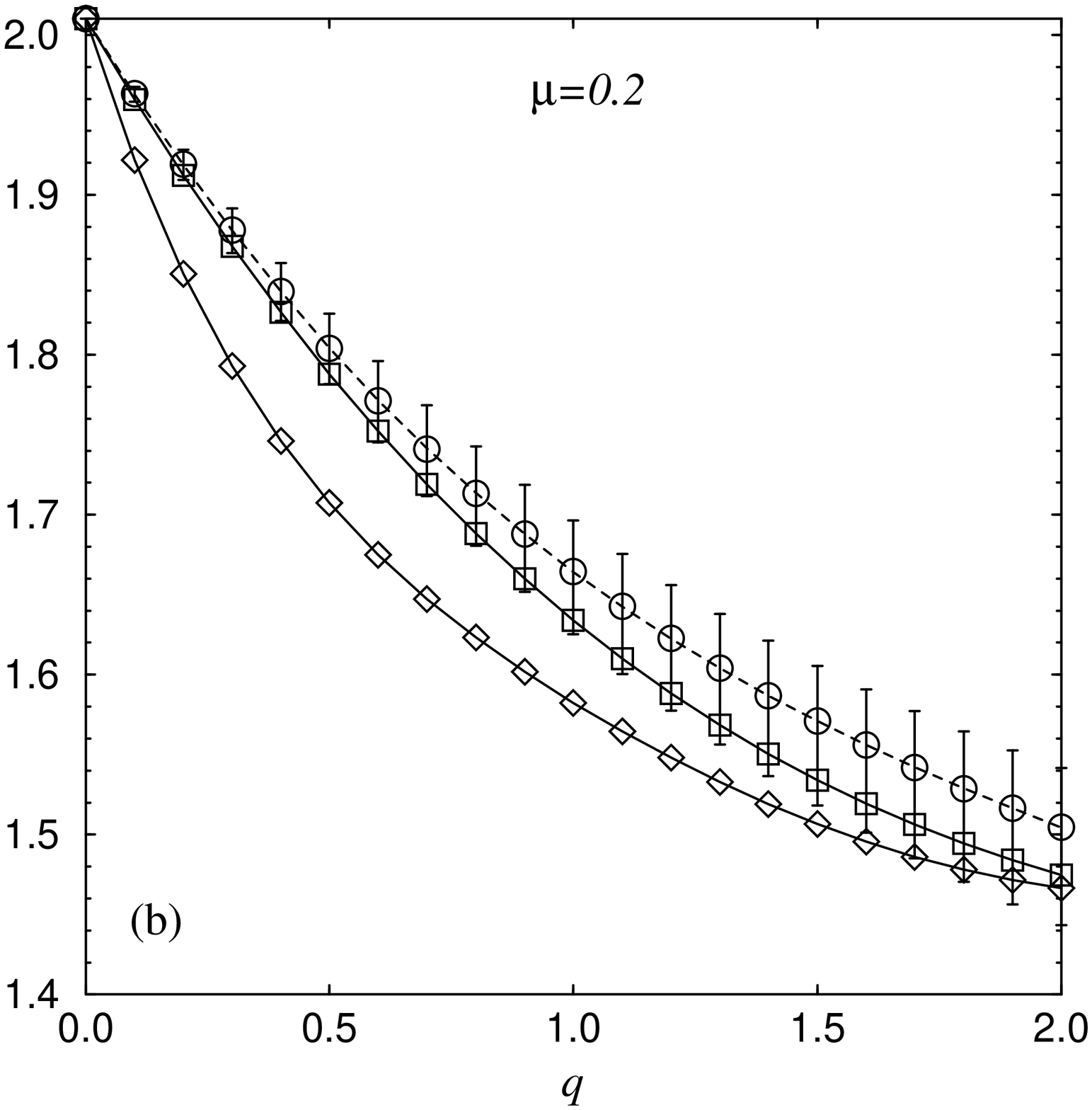}%
\caption{For the case of $\mu=0.2$: (a) $I_1(q,\epsilon)$ for $q$ between
0.1 and 2.0 ($I_2(\epsilon)$ is plotted for $q=1.0$). The dotted lines are linear fits in the
scaling region. (b) $D_q$ computed using numerical solution of \E{weq} (circle with error bar)
and its fourth degree polynomial fit (dotted line). $D_q$ predicted by the theory \E{dqf-2d} when
$\zeta_{2q}$ obtained from numerical simulations are used (square) and when $\zeta_{2q}$ calculated
from \E{zeta2q} are used (diamond).}
\label{dq2}
\end{figure}

To quantify the multifractal nature of $p_i$, we now calculate its R\'{e}nyi
dimension spectrum $D_q$.  We employ the box-counting method to estimate
$D_q$. Using box size $\epsilon/L$ ranging from $2^{-12}$ to $2^{-1}$, we compute
the instantaneous $I_1(q,\epsilon)$ from $t=41$ to $t=75$ at every 1 time unit
using the numerical solution of \E{weq}. For $q \neq 1$, we then make log-log plot
of the time-average of $[I_1/(q-1)]$ versus $\epsilon/L$. These are shown in
\FS{dq1}(a) and \ref{dq2}(a).  For $q=1$, we take the limit $q\rightarrow 1$ in
\E{iqe} to obtain $I_2(\epsilon)\sim \epsilon^{D_1}$, where
\begin{equation}
I_2(\epsilon)=\sum_i p_i(\epsilon)\log_2 [p_i(\epsilon)]\ ,
\label{i2}
\end{equation}
and for $q=1$ in \FS{dq1}(a) and \ref{dq2}(a), the time average of $I_2(\epsilon)$
is plotted against $\log_2(\epsilon/L)$. According to \E{iqe}, these plots will
show a linear region with slope equals $D_q$.  All curves in \FS{dq1}(a) and
\ref{dq2}(a) show slightly undulating behavior which introduces uncertainties in
the determination of $D_q$. The estimated $D_q$ at different values of $q$ are
shown as circles with error bars in \FS{dq1}(b) and \ref{dq2}(b). The error bars
correspond to the variability of the $D_q$ observed at different moments in time.%
\begin{figure}
\includegraphics[scale=0.4]{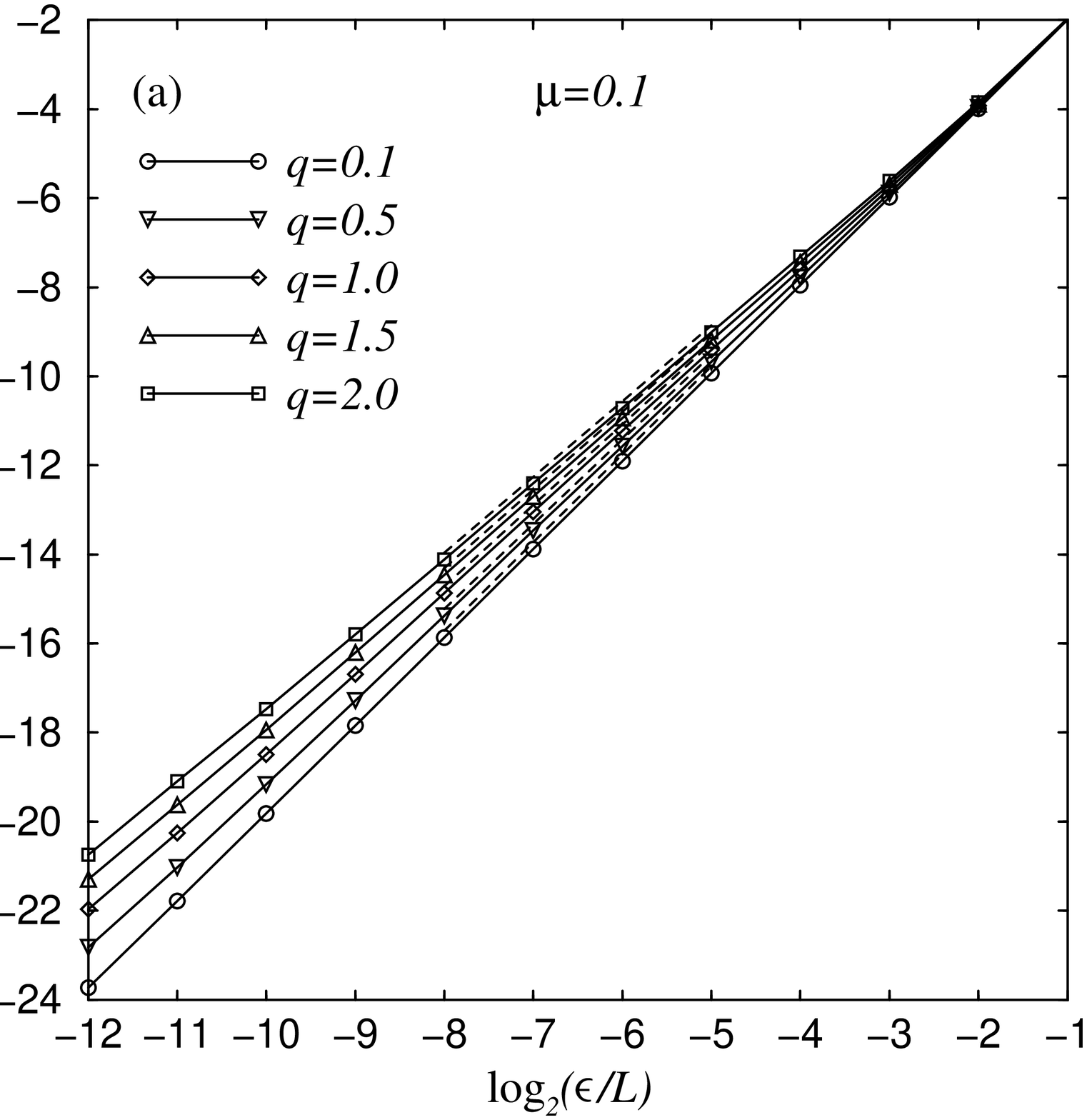}\\%
\vspace{0.6cm}
\includegraphics[scale=0.4]{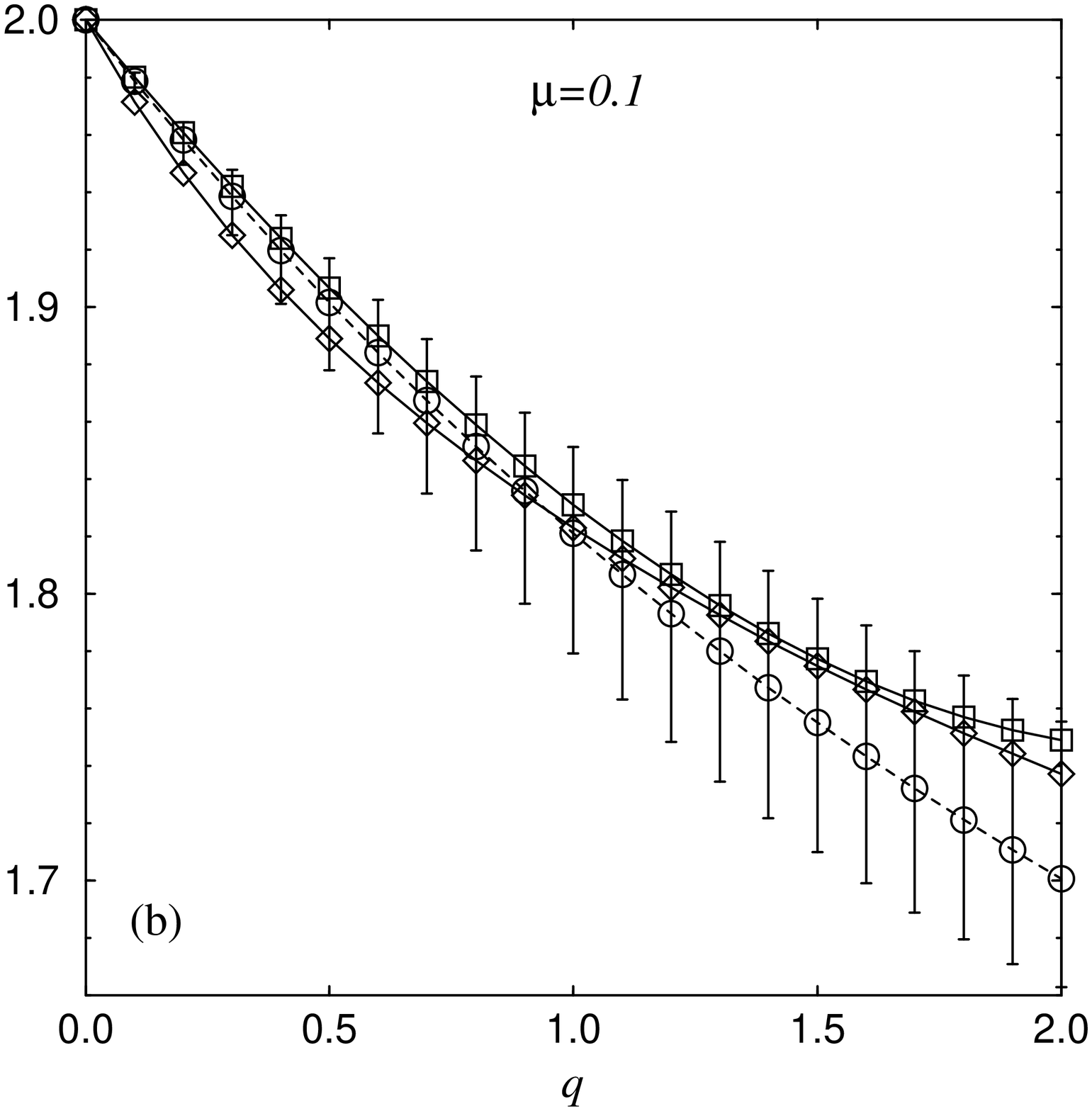}%
\caption{For the case of $\mu=0.1$: (a) $I_1(q,\epsilon)$ for $q$ between
0.1 and 2.0 ($I_2(\epsilon)$ is plotted for $q=1.0$). The dotted lines are linear fits in the
scaling region. (b) $D_q$ computed using numerical solution of \E{weq} (circle with error bar)
and its fourth degree polynomial fit (dotted line). $D_q$ predicted by the theory \E{dqf-2d} when
$\zeta_{2q}$ obtained from numerical simulations are used (square) and when $\zeta_{2q}$ calculated
from \E{zeta2q} are used (diamond).}
\label{dq1}
\end{figure}
The dotted lines in the figures are fourth degree polynomials fitted to the circles. We also
compute $D_q$ using \E{dqf-2d}. To this end, we fit the curves of $\zeta_{2q}/\zeta_2$ versus
$q$ in \F{s1}(b) and \F{s2}(b) with the following polynomial, where $d=3$ for the circles and
$d=5$ for the crosses,
\begin{equation}
\frac{\zeta_{2q}}{\zeta_2}=q\left[1+\sum_{n=1}^{d}a_n(q-1)^n\right]\ .
\label{z2qpoly}
\end{equation}
By \E{dqf-2d}, $D_q$ is then given by
\begin{equation}
D_q=2+\zeta_2\sum_{n=1}^d a_nq(q-1)^{n-1}\ .
\label{dqq}
\end{equation}
In \FS{dq1}(b) and \ref{dq2}(b), we plot \E{dqq} using $\zeta_{2q}$ obtained from numerical
simulations, as well as $\zeta_{2q}$ calculated from our theory. The results are shown as solid lines
labeled with squares and diamonds, respectively. Despite the fact that there are discrepancies
between the $D_q$ obtained by the various methods, they all show the same trend and clearly
indicates that $p_i$ is multifractal ({\it i.e.}, $D_q$ varies with $q$).

We now generate the singularity spectrum $f(\alpha)$ by Legendre transforming the $D_q$
curves shown in \FS{dq1}(b) and \ref{dq2}(b). In particular, for each value of $q$, we have
\cite{halsey86}
\begin{eqnarray}
\alpha(q) &=& \frac{d}{dq}\left[(q-1)D_q\right]\ , \\
f[\alpha(q)] &=& \alpha q - (q-1)D_q\ .
\end{eqnarray}
In the absence of intermittency $D_q$ is independent of $q$ and $f(\alpha)$ is defined only
at $\alpha=D_q$. Thus the consistent determination of $f(\alpha)$ over some range of $\alpha$
indicates intermittency. \F{fa1} and \F{fa2} show the $f(\alpha)$ obtained for $\mu=0.1$ and
$\mu=0.2$ respectively using the $D_q$ obtained by the three methods in \F{dq1}(b) and \ref{dq2}(b)
(circle, square and diamond symbols).%
\begin{figure}
\includegraphics[scale=0.4]{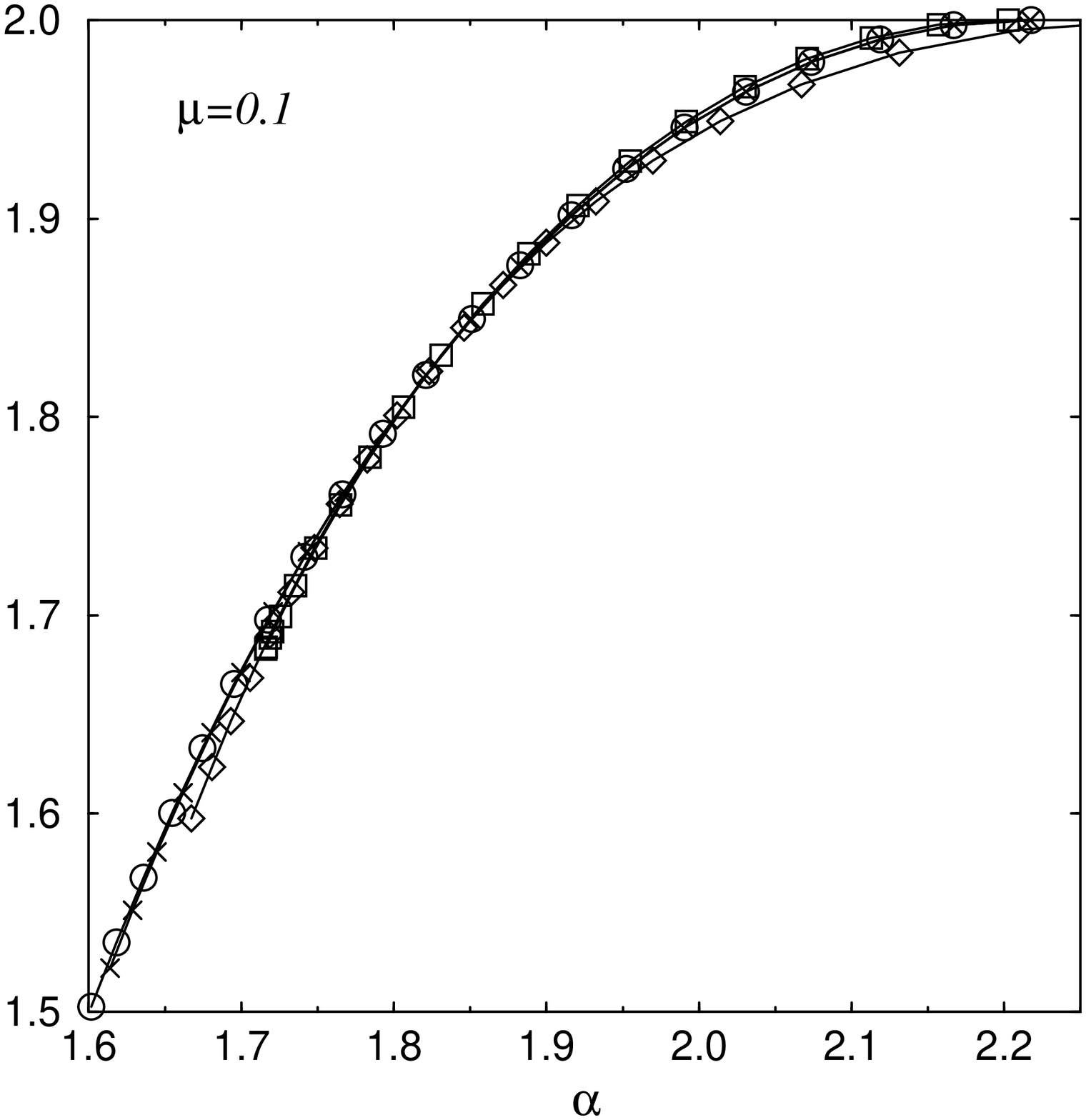}%
\caption{For the case of $\mu=0.1$: The $f(\alpha)$ curves generated by Legendre transforming
the $D_q$ curves in \F{dq1}(b) with the corresponding labels (circle, square and diamond).
$f(\bar{\alpha})$ at different values of $\bar{\alpha}$ obtained by the canonical method is
shown as well (cross).}
\label{fa1}
\end{figure}
\begin{figure}
\includegraphics[scale=0.4]{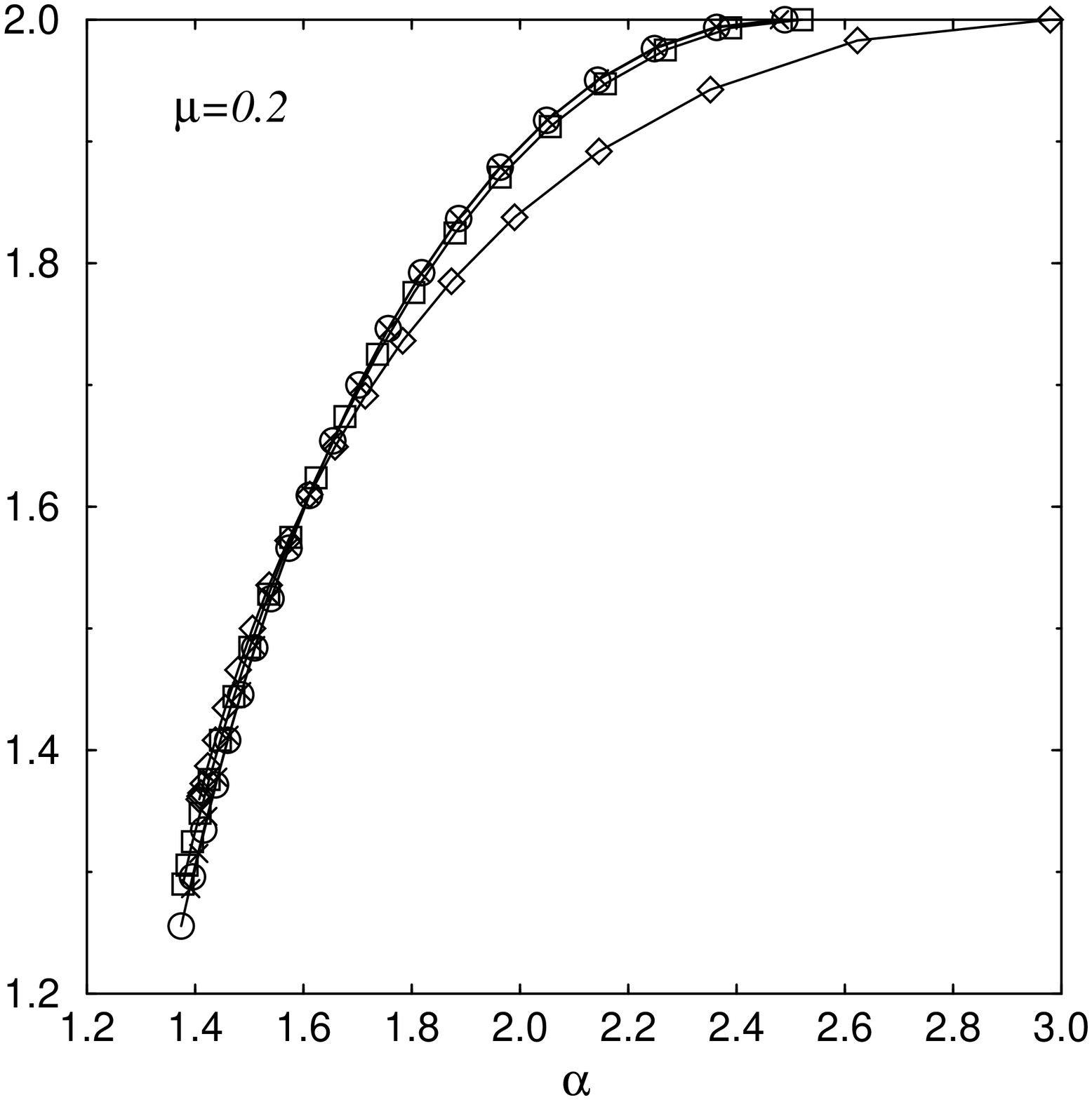}%
\caption{For the case of $\mu=0.2$: The $f(\alpha)$ curves generated by Legendre transforming
the $D_q$ curves in \F{dq2}(b) with the corresponding labels (circle, square and diamond).
$f(\bar{\alpha})$ at different values of $\bar{\alpha}$ obtained by the canonical method is
shown as well (cross).}
\label{fa2}
\end{figure}
The results for $f(\alpha)$ are seen to agree very well with each other in spite of the difference
between the $D_q$ determinations. Thus, we find that $f(\alpha)$ gives a more consistent measure
of intermittency across different methods of determination than does $D_q$. We believe that the
disagreement seen in \FS{dq1}(b) and \ref{dq2}(b) between the different methods for determining
$D_q$ is not significant in view the limited amount of scaling range available.

We can also determine $f(\alpha)$ directly from the numerical solution of \E{weq}. Following
Ref.~\cite{chhabra89b}, we use the canonical method developed in Ref.~\cite{chhabra89a} to
determine $f(\alpha)$. Accordingly, we construct the normalized $q^{\mathrm{th}}$ order measures
$m_i(q,\epsilon)$ in box $\mathcal{R}_i(\epsilon)$ as follow,
\begin{equation}
m_i(q,\epsilon)=\frac{p^q_i(\epsilon)}{\sum_j p^q_j(\epsilon)}\ .
\end{equation}
Then, the mean singularity index $\bar{\alpha}$ and the corresponding $f(\bar{\alpha})$ is
given by
\begin{align}
\bar{\alpha}(q) = \sum_i m_i \alpha_i &= \lim_{\epsilon\rightarrow 0}\lim_{\nu\rightarrow 0}
\frac{\sum_i m_i \log p_i}{\log\epsilon}\ , \\
f[\bar{\alpha}(q)] &= \lim_{\epsilon\rightarrow 0}\lim_{\nu\rightarrow 0}
\frac{\sum_i m_i \log m_i}{\log\epsilon}\ .
\end{align}
One of the advantages of the canonical method is the absence of finite-size effect due to
logarithmic prefactors \cite{chhabra89b}. To estimate $\bar{\alpha}$, we plot the time average
of the quantity $\sum_i m_i \log_2 p_i$ versus $\log_2(\epsilon/L)$ and measure the slopes in
the scaling region. Similarly, when we plot the time average of $\sum_i m_i \log_2 m_i$ versus
$\log_2(\epsilon/L)$, the slopes in the scaling range give the values of $f(\bar{\alpha})$. For
the case $q=1$, $\bar{\alpha}(1)=f[\bar{\alpha}(1)]$ is obtained by measuring the slope of the
curve of the time-averaged $I_2(1,\epsilon)$ versus $\log_2(\epsilon/L)$. The values of
$f(\bar{\alpha})$ at different values of $\bar{\alpha}$ obtained by this scheme are shown as
crosses in \F{fa1} and \F{fa2} for $\mu=0.1$ and $\mu=0.2$ respectively. They are in excellent
agreement with the $f(\alpha)$ generated by Legendre transforming the $D_q$ curve obtained from
the numerical solution of \E{weq}.

\subsection{Discussion}

Kolmogorov introduced the refined similarity hypothesis (RSH) \E{rsh} to take into account the spatial
fluctuations of $\varepsilon$ in three-dimensional turbulence. The relation \E{dqf-3d} is a direct
consequence of the RSH. In two-dimensional turbulence, the relevant quantity is the local rate of
viscous enstrophy dissipation $\eta$. We have already seen that the measure $p_i$ is multifractal
in the presence of drag, indicating the intermittent nature of $\eta$. Following Kolmogorov's ideas,
we consider the average of $\eta$ over an area $\mathcal{R}(r)$ of linear dimension $r$:
\begin{equation}
\eta_r=\frac{\nu}{r^2}\int_{\mathcal{R}(r)}|\nabla\omega|^2\,d\V{x}\ .
\end{equation}
Analogous to the Kolmogorov's RSH, we propose that at high Reynolds number, there is an
inertial range of $r$ in which
\begin{equation}
\avg{|\delta_{\V{r}}\omega|^{2q}\ \vline\ \eta_r} = C_{2q}(r^{\zeta_2}\eta_r)^q
\label{rsh2d}
\end{equation}
where $C_{2q}$ are constants. \E{rsh2d} implies $\avg{|\delta_{\V{r}}\omega|^{2q}}
\sim\avg{(\eta_r)^q}r^{q\zeta_2}$ and hence $\avg{(\eta_r)^q}\sim r^{\gamma_q}$ with
\begin{equation}
\gamma_q=\zeta_{2q}-q\zeta_2\ .
\label{gamma}
\end{equation}
An expression analogous to \E{gamma} has been proven for Kraichnan's model of passive
scalar advection \cite{kraichnan95,chertkov96}. We note that \E{rsh2d} follows if
\begin{equation}
\eta_r \sim \frac{(\delta_{\V{r}}\omega)^2}{r^{\zeta_2}}
\end{equation}
which is the two-dimensional counterpart of \E{er}. It is straightforward to show that the
hypothesis \E{rsh2d} implies \E{dqf-2d}. Denoting $\eta_r$ in box $\mathcal{R}_i(r)$ by $\eta_r^{(i)}$,
we have
\begin{equation}
p_i(r)=\frac{r^2}{L^2}\frac{\eta_r^{(i)}}{\avg{\eta}}\ .
\end{equation}
Since $\sum_i[\eta_r^{(i)}]^q \sim r^{-2}\avg{(\eta_r)^q}$, we get $I(q,r)\sim r^{2q-2+\gamma_q}$,
from which \E{dqf-2d} immediately follows.

Finally, we remark that if the measure $p_i$ is defined with a parameter $n$ as follow \cite{varosi91},
\begin{equation}
p_i(\epsilon,n)
=\frac{\int_{\mathcal{R}_i(\epsilon)}|\nabla\omega|^n\,d\V{x}}
{\int_\mathcal{R}|\nabla\omega|^n\,d\V{x}}\ ,
\end{equation}
then the corresponding formula for the $D_q$ based on this measure is
\begin{equation}
D_q=2+\frac{\zeta_{nq}-q\zeta_n}{q-1}\ .
\end{equation}

\section{Conclusion}

We have studied the enstrophy cascade regime of two-dimensional turbulence with
linear drag.  A previous theory for the power law exponent of the energy
wavenumber spectrum is verified by direct numerical computation using a
$4096\times 4096$ lattice. We also calculate the vorticity structure functions
numerically and show that they exhibit anomalous scaling in the presence of
drag. The values of the structure function scaling exponents $\zeta_{2q}$ are
measured and found to agree with the prediction by previous theory.  We then
compute the probability distribution function $\bar{P}_r(X_{\V{r}})$ of the
standardized vorticity difference $X_{\V{r}}$ and find that $\bar{P}_r(X_{\V{r}})$
develops exponential and stretched-exponential tails at small values of $r$. The
theoretical expression for $\bar{P}_r(X_{\V{r}})$ is shown to give predictions
that agree well with the numerical results for a wide range of $r$. A measure
based on the local viscous enstrophy dissipation rate $\eta$ is studied in terms
of its R\'{e}nyi dimension spectrum $D_q$ and singularity spectrum $f(\alpha)$,
and is found to be multifractal.  The intermittency in $\eta$ is connected to the
intermittency in vorticity difference by a two-dimensional analog of the refined
similarity hypothesis, and we derive a formula that relates $D_q$ to $\zeta_{2q}$.

\begin{acknowledgments}

This work was supported by the Office of Naval Research and by the National Science Foundation
(PHYS 0098632).

\end{acknowledgments}

\bibliography{vort_arxiv}

\end{document}